\documentclass[useAMS,usenatbib,referee]{biom}
\def\bSig\mathbf{\Sigma}

\usepackage{graphicx}
\usepackage{amsfonts}
\usepackage{amsmath}
\usepackage[shortlabels]{enumitem}
\usepackage{caption, subcaption}

\newcommand{\bigCI}{\mathrel{\text{\scalebox{1.07}{$\perp\mkern-10mu\perp$}}}}

\newtheorem{Corollary}{Corollary}

\title[Valid post-selection inference for hazard ratios]{Ensuring valid inference for hazard ratios after variable selection}

\author
{Kelly Van Lancker\emailx{kelly.vanlancker@ugent.be} \\
Department of Biostatistics, Johns Hopkins Bloomberg School of Public Health, Baltimore, U.S.A.\\
Department of Applied Mathematics, Computer Science and Statistics, Ghent University, Ghent, Belgium
\and
Oliver Dukes\emailx{oliver.dukes@ugent.be} \\
Department of Statistics and Data Science, The Wharton School, University of Pennsylvania, Philadelphia, U.S.A.\\
Department of Applied Mathematics, Computer Science and Statistics, Ghent University, Ghent, Belgium
\and
Stijn Vansteelandt\emailx{stijn.vansteelandt@ugent.be} \\
Department of Applied Mathematics, Computer Science and Statistics, Ghent University, Ghent, Belgium\\
Department of Medical Statistics, London School of Hygiene and Tropical Medicine, London, U.K.}

\begin{document}





\pagerange{\pageref{firstpage}--\pageref{lastpage}} 


\doi{10.1111/j.1541-0420.2005.00454.x}


\label{firstpage}


\begin{abstract}
The problem of how to best select variables for confounding adjustment forms one of the key challenges in the evaluation of exposure effects in observational studies, and has been the subject of vigorous recent activity in causal inference. A major drawback of routine procedures is that there is no finite sample size at which they are guaranteed to deliver exposure effect estimators and associated confidence intervals with adequate performance.  
In this work, we will consider this problem when inferring conditional causal hazard ratios from observational studies under the assumption of no unmeasured confounding. The major complication that we face with survival data is that the key confounding variables may not be those that explain the censoring mechanism. 
In this paper, we overcome this problem using a novel and simple procedure that can be implemented using off-the-shelf software for penalized Cox regression. In particular, we will propose tests of the null hypothesis that the exposure has no effect on the considered survival endpoint, which are uniformly valid under standard sparsity conditions.
Simulation results show that the proposed methods yield valid inferences even when covariates are high-dimensional.
\end{abstract}

%



\maketitle


%

\section{Introduction}
The effect of an exposure on a time-to-event endpoint in the presence of right censoring is routinely estimated as the exposure coefficient in a Cox proportional hazards model \citep{Cox1972}, adjusted for baseline covariates. 
However, it is rarely known \textit{a priori} which covariates to include in order to render the adjustment sufficient to control for confounding and non-informative censoring.
It is likewise unknown how to correctly specify the functional form with which these covariates should enter the model. Data-adaptive procedures (e.g., variable selection procedures) are therefore typically employed in order to select which variables to adjust for and how. In particular, they become inevitable when the number of covariates is close to or greater than the sample size.

Routine practice is often based on stepwise selection procedures that use hypothesis testing or regularization techniques like Lasso regression \citep{tibshirani1997}. Since confounders are by definition associated with outcome, these procedures may succeed well in detecting strong confounders, in addition to variables that are purely predictive of survival. 
However, they may fail to detect confounders that are important because they are strongly associated with exposure while only moderately associated with outcome. This results in improper adjustment for confounding, which can in turn translate into bias. While this is true for any type of regression analysis \citep{leeb2006}, survival analyses suffer additional complications due to censoring. 
In particular, failing to control for prognostic factors of outcome as a result of variable selection errors may induce informative censoring bias when those covariates are also associated with censoring. 
Such selection errors are likely to occur in variables that have a weak-to-moderate effect on the outcome but a strong effect on censoring.
These problems persist with increasing sample size as one can always find data-generating mechanisms at which an imperfect selection is made. There thus exists no finite sample size $n$ at which normal-based tests and intervals are guaranteed to perform well.

In this article, we will provide two strategies for obtaining valid post-selection inference w.r.t. the exposure effect parameter indexing a Cox model. The first is simple and intuitive, whilst the second has a sharper justification based on recent theory for high-dimensional inference. 
In particular, we will rely on the selection of variables associated with either survival, censoring or exposure via three separate selection steps followed by a final estimation step. Our first proposal is as follows:
\begin{enumerate}
	\item  We first select variables that predict outcome by fitting a standard Cox model for the survival time given exposure and baseline covariates using the Lasso. This step helps to pick up important confounders, control variables for censoring adjustment and variables for which adjustment may increase power of the test of no exposure effect. 
	\item In the second step, we select variables that predict censoring by fitting a Cox model for (the cause-specific hazard of) censoring given exposure and baseline covariates using the Lasso. This step provides a second chance to pick up variables that may explain censoring, which is especially relevant when those variables are only weakly-to-moderately predictive of survival.
	\item In the third step, we select variables that predict exposure by fitting a model (e.g., linear or logistic) for exposure given baseline covariates using the Lasso. This step provides a second chance to pick up confounders, especially those that are strongly related to exposure (and possibly only weakly-to-moderately predictive of survival). This step is redundant when the exposure is randomly assigned.
	\item Finally, we estimate the exposure effect of interest by fitting a Cox model for the survival time given exposure and the union of the sets of variables selected in the three variable selection steps.
\end{enumerate}
This proposed ``poor man's approach'' extends earlier work on randomized experiments \citep[see][]{vanlancker2020} to observational data. 
The second proposal that we will make in this paper is more rigorous and builds on the theory of high-dimensional inference \citep[see][]{bradic2011, huang2013}.
In particular, it draws on the ``double selection'' approach for generalized linear models,  proposed by \cite{belloni2014, belloni2016} to develop uniformly valid post-selection inference \citep[see also][]{vandegeer2014, zhangzhang2014, ning2017}. Double
selection was developed in the context of selection of confounders by using two steps to identify covariates for inclusion: first selecting variables that predict outcome and then those
that predict exposure.
We develop a ``triple selection'' approach that extends the work of \citet{belloni2014} to survival data. The corresponding test and estimator are related to a recent proposal by \citet{fang2017}, but are found to perform better under most censoring mechanisms in extensive Monte-Carlo simulation studies and have the added advantage of being obtainable using standard statistical software.


At this point we note that the causal interpretation of the hazard ratio is subtle, even in the absence of model misspecification and unmeasured confounding \citep{hernan2010}. 
This is because the hazard ratio has a built-in selection bias by conditioning on survival up to a given time point; even if exposed and unexposed patients are comparable at baseline, this may no longer be true at later time points.
Nonetheless, the use of
the Cox model is widespread partly because the hazard ratio serves as a convenient measure of association which may often be constant  in time and covariates to a reasonable degree of approximation. Indeed, the hazard ratio is often the main and only effect measure reported in epidemiologic studies \citep[e.g.,][]{hernan2010, uno2014moving}. It is therefore important to give advice and recommendations to practitioners on how to conduct these analyses. Moreover, tests of the causal null hypothesis (which form much of the focus of this paper) obtained from the Cox model are not vulnerable to the aforementioned selection bias.

The rest of this paper is organized as follows. In Section \ref{sec:motivation}, we state the null hypothesis of interest and describe the challenges with obtaining valid inference after variable selection. 
In Section \ref{sec:proposal}, we propose a simple, heuristic as well as a more rigorous approach for testing hypotheses and obtaining confidence intervals. For the latter approach, we give asymptotic guarantees regarding type I error and interval coverage.
We investigate the empirical performance of these methods in Section \ref{sec:simulation}. In Section \ref{sec:dataAnalysis}, we illustrate the proposal on the Breast cancer data set used in \cite{royston2013external}. We end with a discussion in Section \ref{sec:discussion}.

\section{Motivation}\label{sec:motivation}
We begin with some notation. Let $T$ be the survival time and $C$ the censoring time. We observe $\left(U, \delta\right),$ where $U=\min (T, C)$ is the
observed portion of $T$ and $\delta=I(T \leqslant C)$ is a censoring indicator. Let $L=(L_1, \dots, L_p)'$ be the $p$-dimensional vector of baseline covariates and $A$ the exposure. 
The observed data is then given by the i.i.d. sample $\{(U_i, \delta_i, L_i, A_i), i=1, \dots, n\}$. Let $\tau$ denote the end-of-study time.
Throughout, we assume that censoring is non-informative conditional on $A$ and $L$ in the sense that $T \bigCI C |A, L$ and that $L$ is sufficient to adjust for confounding of the effect of $A$ on $T$. 
Suppose for the moment that we are interested in testing the null hypothesis that the event time distribution does not depend on the exposure $A$ conditional on $L$. 

We assume that the conditional hazard rate function at time $t$, $\lambda \left(t | A, L\right)$, obeys the proportional hazards model \citep{Cox1972}
\[\lambda\left(t\mid A, L\right)=\lambda_0(t, \alpha^*, \beta^*)e^{\alpha^{*} A+\beta^{*'}L}=\lambda_0(t)e^{\alpha^{*} A+\beta^{*'}L},\]
where $\lambda_{0}(t)$ is an unspecified baseline hazard function, $\alpha^*$ encodes the unknown exposure effect of interest and also $\beta^*\in\mathbb{R}^{p}$ is an unknown parameter.
In the classical low dimensional setting (with fixed $p$), one may exploit the profile partial score function
\begin{align}\label{eq:score_low}
	\qquad S(\alpha, \beta)
	&=-\frac{1}{n} \sum_{i=1}^{n} \int_{0}^{\tau}\left[A_{i}-\frac{\mathbb{E}_n\left\{ AR(t) e^{\alpha A+\beta'L}\right\}}{\mathbb{E}_n\left\{ R(t) e^{\alpha A+\beta'L }\right\}}\right] \mathrm{d} N_{i}(t),
\end{align}
to test the null hypothesis that $\alpha^*=0$, with $\mathbb{E}_n\left\{\cdot\right\}$ refering to the sample average. Here, $R_i(t)$ is the at risk indicator (which is the product of the indicators $I(T_i\geq t)$ and $I(C_i\geq t)$) and $dN_i(t)$ the increment in the counting process with respect to the event time $T_i$, $N_i(t):=I(U_i\leq t, \delta_i=1)$, at time $t$ for subject $i$ ($i\in 1, \dots, n$).
Define $\hat{\beta}(\alpha)$ as the maximum partial likelihood estimator for $\beta^*$ at a fixed $\alpha$ (i.e., fixed at zero to test $\alpha^*=0$). 
We can then obtain an asymptotically unbiased and normal score test statistic based on the (properly scaled) profile partial score function $S\left(\alpha, \hat{\beta}(\alpha)\right)=S\left(0, \hat{\beta}(0)\right)$.

However, in practice, there is often little prior knowledge on which variables in a given dataset require adjustment to control for confounding or non-informative censoring, as well as how to model the association between these variables and outcome. Hence, data-adaptive procedures are typically employed in order to select the variables to adjust for. 
The standard methodology described above does not straightforwardly
extend to settings where variable selection is employed.
To see this, let $\tilde{\beta}$ denote
an estimator of $\beta^*$ obtained either directly via some regularization method or
after model selection with $\alpha$ fixed at zero. Define
\begin{align*}
	\mathcal{U}_{i}(\tilde{\beta})= \int_{0}^{\tau}\left[A_{i}-\frac{\mathbb{E}_n\left\{  AR(t) e^{\tilde{\beta}'L}\right\}}{\mathbb{E}_n\left\{ R(t) e^{\tilde{\beta}'L}\right\}}\right] \mathrm{d} N_{i}(t).
\end{align*}
The Taylor expansion,
\begin{align} \label{TaylorExp}
	\begin{split}
		\frac{1}{\sqrt{n}} \sum_{i=1}^{n} \mathcal{U}_{i}(\tilde{\beta})=& \frac{1}{\sqrt{n}} \sum_{i=1}^{n} \mathcal{U}_{i}(\beta^*)+\frac{1}{n} \sum_{i=1}^{n} \frac{\partial \mathcal{U}_{i}(\beta^*)}{\partial \beta} \sqrt{n}(\tilde{\beta}-\beta^*)+O_{P}\left(\sqrt{n}\|\tilde{\beta}-\beta^*\|_{2}^{2}\right),
	\end{split}
\end{align}
where $\|\cdot\|_{2}^{2}$ denotes the Euclidian norm, then provides insight into the asymptotic distribution of $\tilde{\beta}$. For fixed $\beta^*$ the remainder term $O_{P}\left(\sqrt{n}\|\tilde{\beta}-\beta^*\|_{2}^{2}\right)$ converges to zero when $\tilde{\beta}$ converges sufficiently quickly. 
However, this pointwise result does not reflect performance over the parameter space in finite samples. In particular, it does not prevent the existence of converging sequences $\beta_n$ for which
$\sqrt{n}(\tilde{\beta}-\beta_n)$ (and thus the test statistic) has a complex, non-normal distribution which is (possibly) centered away from zero. This is a result of the discrete nature of data-adaptive methods which force $\tilde{\beta}$ to zero in some samples (but not in others). 
Through the second term on the righthand side of the above equality, this complex distribution may then propagate into a complex distribution of the test statistic.

At a more intuitive level, we can understand the bias that arises via variable selection by mistakenly removing confounders or variables that render censoring non-informative.
First, standard approaches fail by missing important confounders that are weakly predictive of outcome, but strongly associated with exposure. This is because the selection procedure may lack power due to collinearity between exposure and confounders. 
We may therefore fail to properly adjust for confounding, which can translate into bias. Survival analyses unfortunately face additional challenges due to censoring. 
In particular, failing to control for certain baseline covariates as a result of variable selection errors may induce informative censoring. 
Selection mistakes are likely to occur for variables that have a moderate effect on the outcome but a strong effect on censoring.
Selection strategies may lack power to detect such variables because censoring implies information loss and may even reduce the variation in certain baseline variables in the risk set.
Imperfect selection may lead to collider-bias (sometimes referred to as selection bias, sampling bias, ascertainment bias) between $A$ and $T$ in the risk set at each time point, which can in turn induce a distorted association where there is no effect in the general population. 
This is even problematic in randomized trials, whose analysis is immune to confounding bias \citep[see][]{vanlancker2020}.

As previously suggested, this problem persists with increasing sample size as for each sample size $n$ one can find values of $\beta$ at which imperfect selection happens with high probability. The score test statistic therefore does not converge uniformly over the parameter space to the limiting standard normal distribution \citep{leeb2006}. There is then no guarantee that the procedure will work well in finite samples as there is no finite $n$ at which the normal approximation is guaranteed to hold.

\section{Proposal for Variable Selection}\label{sec:proposal}
Building on the work of \cite{belloni2014, belloni2016} and \cite{vanlancker2020}, we will first make a simple proposal to overcome the previous problems. Next, we will compare this proposal with a more rigorous but complex approach.
\vspace{-0.5cm}
\subsection{Poor Man's Approach}\label{sec:poor}
In order to achieve valid post-selection inference, we recommend a ``triple selection'' approach based on the Lasso \citep{tibshirani1997}.
Inspired by the double selection approach \citep{belloni2014, belloni2016}, our proposal will rely on the selection of variables associated with either survival, censoring or exposure via three separate models. By using three different variable selection steps followed by a final estimation step, we aim to overcome the problem of standard approaches that rely on a single selection step (see Section \ref{sec:motivation}). 
In particular, this approach makes it more likely to detect variables that demand adjustment but are otherwise difficult to diagnose because they are strongly associated with exposure and/or censoring, while having only a small-to-moderate effect on the outcome.

In the remainder of the paper, we will focus on the Lasso since it is known to perform well with a large number of covariates and is readily available in many statistical software packages. However, our proposal can allow for more general variable selection procedures.
Building on the idea of double selection, we perform a three stage selection procedure followed by a final
estimation step as follows:
\begin{enumerate}
	\item Fit a Cox model for the hazard $\lambda\left(t\mid A, L\right)$ corresponding with the survival time $T$ given exposure $A$ and baseline covariates $L$ using the Lasso (penalizing all coefficients in the model, including the exposure), $\lambda\left(t\mid A, L\right)=\lambda_0(t, \alpha^*, \beta^*)e^{\alpha^{*} A+\beta^{*'}L}$,
	and select the covariates with non-zero estimated coefficients.
	\item Fit a Cox model for the hazard $\lambda^C\left(t\mid A, L\right)$ corresponding with the time to censoring $C$ given exposure $A$ and baseline covariates $L$ using the Lasso (penalizing all coefficients in the model, including the exposure), $\lambda^C\left(t\mid A, L\right)=\lambda_0^C(t)e^{\eta_1^{*} A+\eta_2^{*'}L}$
	(viewing the times at which events occur as censored), where $\lambda_0^C(t)$ is an unknown baseline hazard,
	and select the covariates with non-zero estimated coefficients.
	\item Fit a model (e.g., linear or logistic) for exposure $A$ on baseline covariates $L$ using the Lasso (penalizing all coefficients in the model), $E\left(A\mid L\right)=g^{-1}\left(\delta_0^{*}+\delta_1^{*'}L\right)$,
	with $g$ a known link function (e.g., identity for linear models and logit for logistic models), and select the covariates with non-zero estimated coefficients.
	\item  
	Fit a Cox model for the survival time $T$ given exposure $A$ and all covariates selected in either one of the first three steps to obtain final estimates $\hat{\alpha}_{PM}$ and $\hat{\beta}_{PM}$ for $\alpha^*$ and $\beta^*$. 
	This regression may also include additional variables that were not selected in the first three steps, but that were identified \textit{a priori} as being important.\vspace{-0.25cm}
\end{enumerate}

Inference on the treatment effect $\hat{\alpha}_{PM}$ may then be performed using conventional methods, provided that a robust standard error is used. This can be intuitively understood upon noting that our procedure will only reject covariates that are weakly associated
with survival, censoring and exposure; asymptotically, this is not problematic because the omission of such covariates induces such weak degrees of informative censoring and/or confounding that the resulting bias in the test statistic for the null will be small enough that inference is not jeopardised. 

In the next section, we will introduce a related proposal, which is more complex but whose validity is easier to understand. This proposal will give justification to the poor man's approach.


\subsection{Triple Selection}\label{sec:debiased}
Motivated by the problem described in Section \ref{sec:motivation}, we will make use of a different test statistic/estimator for $\alpha$ (than that based on the profile partial score function) that is asymptotically normal under standard conditions, even in conjunction with high dimensional variable selection. 
The method proposed here differs from the poor man's approach in how the predictors of the exposure are selected (Step $3$ in Section \ref{sec:poor}). In particular, we will fit the exposure model in a specific way which de-biases the na\"ive estimator of the exposure effect so as to obtain valid inference:
\begin{enumerate}
	\item Fit a Cox model for survival time $T$ given exposure $A$ and baseline covariates $L$ using the Lasso (penalizing all coefficients in the model), 
	select the covariates with non-zero estimated coefficients,
	and refit the Cox model on exposure $A$ and all covariates selected by the Lasso step to obtain estimates $\hat{\alpha}$ and $\hat{\beta}$ (where the components of $\hat{\beta}$ with non-selected variables are set to zero). 
	\item Fit a Cox model for (the cause-specific hazard of) censoring $C$ on exposure $A$ and baseline covariates $L$ using Lasso (penalizing all coefficients in the model), 
	and select the covariates with non-zero estimated coefficients.
	\item Fit a linear model for the Schoenfeld residuals for the exposure
	$ A_i-\bar{A}_n(T_i, \hat{\alpha}, \hat{\beta}),$
	where $\bar{A}_n(t, \alpha, \beta)=\frac{ \mathbb{E}_n\left\{ AR(t) e^{\alpha A+\beta'L}\right\}}{\mathbb{E}_n\left\{ R(t) e^{\alpha A+\beta'L}\right\}}$, on the Schoenfeld residuals for the covariates
	$ L_i-\bar{L}_n(T_i, \hat{\alpha}, \hat{\beta}),
	$
	where $\bar{L}_n(t, \alpha, \beta)=\frac{ \mathbb{E}_n\left\{ LR(t) e^{\alpha A+\beta'L}\right\}}{\mathbb{E}_n\left\{ R(t) e^{\alpha A+\beta'L}\right\}}$, in subjects for whom an event was observed ($\delta_i=1$) at the corresponding event time $T_i$ using the Lasso (penalizing all coefficients in the model), and select the covariates with non-zero estimated coefficients.
	Here, $\hat{\alpha}$ and $\hat{\beta}$ are the post-Lasso estimates obtained in Step 1.
	\item  Fit a Cox model for the survival time $T$ given exposure $A$ and all covariates selected in either one of the first three steps to obtain final estimates $\check{\alpha}$ and $\check{\beta}$ for $\alpha^*$ and $\beta^*$.
	This regression may additionally include a small set of additional covariates identified \textit{a priori} as
	necessary. 
	Inference on the treatment effect $\check{\alpha}$ may then be performed using conventional methods, provided that a robust standard error is used (see Part 2 of the proof of Theorem \ref{th:main} in Appendix A of the online supplementary materials for justification). \vspace{-0.25cm}	
\end{enumerate}
Because of its link with the groundbreaking work of \cite{belloni2014, belloni2016}, we will refer to this method as the (post)-triple selection approach.
\subsubsection{Intuition for the Importance of Triple Selection}\label{sec:intuition}
The method proposed in the previous section overcomes the problem described in Section \ref{sec:motivation} by using a different score test statistic for $\alpha$ that is asymptotically normal under standard conditions, even when high dimensional variable selection is used.
Key to the choice of score is that it decorrelates the score function of the primary parameter ($\alpha$) from that of the nuisance parameters ($\beta$). 
Our analysis is therefore based on the decorrelated score function that takes into account the estimation of the nuisance parameters,
\begin{align}\label{eq:scoreFunctExp}
\begin{split}
	U_{i}\left(\alpha, \beta, \gamma\right)=\int_0^\tau\left[A_{i}-\bar{A}(t, \alpha, \beta)-\gamma'\left\{
	L_{i}-\bar{L}(t, \alpha, \beta)
	\right\} \right]\\
	\times\left\{dN_i(t)-\lambda_{0}(t, \alpha, \beta)e^{\alpha A_i+\beta' L_i}R_i(t)dt\right\},
\end{split}
\end{align}
with 
\[\bar{A}(t, \alpha, \beta)=\frac{\mathbb{E}\left\{AR(t) e^{\alpha A+\beta'L}\right\}}{\mathbb{E}\left\{R(t) e^{\alpha A+\beta'L}\right\}}\quad \textrm{and}\quad \bar{L}(t, \alpha, \beta)=\frac{\mathbb{E}\left\{LR(t) e^{\alpha A+\beta'L}\right\}}{\mathbb{E}\left\{R(t) e^{\alpha A+\beta'L}\right\}}, \]
and where $\gamma$ is a $p$-dimensional parameter with population value $\gamma^*$ defined as the population ordinary least squares coefficient from a least squares regression of the exposure Schoenfeld residuals $A_{i}-\bar{A}(T_i, \alpha^*, \beta^*)$ on the covariate Schoenfeld residuals $L_{i}-\bar{L}(T_i, \alpha^*, \beta^*)$ in subjects with an event. 
Note that, in high dimensions, we estimate $\gamma^*$ via the Lasso (see Step $3$) and define $\hat{\gamma}$ as the corresponding post-Lasso estimator obtained in Step 3 of the triple selection approach. 
Let $B$ denote the (index of the) variables selected in Step 1, 2 and 3 of the triple selection approach.
We then define $\check{\gamma}$ as the estimate of $\gamma^*$ obtained via a ordinary least squares regression of $A_{i}-\bar{A}_n(T_i, \check{\alpha}, \check{\beta})$ on $L_{i}-\bar{L}_n(T_i, \check{\alpha}, \check{\beta})$ in subjects with an event and subject to $\left\{j\in\{1, \dots, p\}: \check{\gamma}_j\neq 0\right\}\subseteq B$ (i.e., $\check{\gamma}_j=0$ for $j\notin B$). 




In what follows, we focus our developments on the function
\begin{align*}
  \hat{U}_{i}\left(\alpha, \beta, \gamma\right)=\int_0^\tau\left[A_{i}-\bar{A}_n(t, \alpha, \beta)-\gamma'\left\{
L_{i}-\bar{L}_n(t, \alpha, \beta)
\right\} \right]\{dN_i(t)-R_i(t)\hat{\lambda}_0(t, \alpha, \beta)e^{\alpha A_i+\beta' L_i}dt\},  
\end{align*}
where $\hat{\lambda}_0(t, \alpha, \beta)=\frac{\mathbb{E}_n\left\{R(t)dN(t)\right\}}{\mathbb{E}_n\left\{R(t)e^{\alpha A+\beta' L}\right\}}$ and where we substitute the population averages $\bar{A}(t, \alpha, \beta)$ and $\bar{L}(t, \alpha, \beta)$ in Expression \eqref{eq:scoreFunctExp} with $\bar{A}_n(t, \alpha, \beta)$ and $\bar{L}_n(t, \alpha, \beta)$.
Via a Taylor expansion of $\frac{1}{\sqrt{n}} \sum_{i=1}^{n} \hat{U}_{i}(\check{\alpha}, \check{\beta}, \check{\gamma})$ around $(\alpha^*, \beta^*, \gamma^*)$, we obtain
\begin{align}
	\begin{split}
		\sqrt{n}(\check{\alpha}-\alpha^*)&=-\left\{\frac{1}{\sqrt{n}} \sum_{i=1}^{n} U_{i}(\alpha^*, \beta^*, \gamma^*)\right\}V^{*-1} \\
		&-\sqrt{n}(\check{\beta}-\beta^*)'\left\{\frac{1}{n} \sum_{i=1}^{n} \frac{\partial}{\partial \beta} U_{i}(\alpha^*, \beta^*, \gamma^*)\right\}V^{*-1}\label{eq:taylor_alpha}\\
		&-\sqrt{n}(\check{\gamma}-\gamma^*)'\left\{\frac{1}{n} \sum_{i=1}^{n} \frac{\partial}{\partial \gamma} U_{i}(\alpha^*, \beta^*, \gamma^*)\right\}V^{*-1}+\text{ Remainder},
	\end{split}
\end{align}
where the remainder contains second order terms and $V^*=\mathbb{E}\left\{\frac{\partial}{\partial \alpha} U_{i}(\alpha^*, \beta^*, \gamma^*)\right\}$ is a scaling factor. By ensuring that $n^{-1}\sum_{i=1}^n\partial U_i(\alpha^*, \beta^*, \gamma^*)/\partial\gamma$ and $n^{-1}\sum_{i=1}^n\partial U_i(\alpha^*, \beta^*, \gamma^*)/\partial\beta$ converge to zero in probability, we guarantee that the complex, non-standard behavior of the estimators $\check{\gamma}$ and $\check{\beta}$ does not affect the asymptotic behaviour of the test for $\alpha^*$ nor the estimator $\check{\alpha}$. 
Specifically, for the gradient with respect to $\gamma$ this is achieved by solving the following penalized estimating equation for $\theta=(\alpha, \beta)$
\begin{align}
	\begin{split}\label{eq:subgrad_beta}
		0&=
		\frac{1}{n}\sum_{i=1}^n\frac{\partial \hat{U}_i(\alpha,\beta, \gamma)}{\partial\gamma}+\lambda_{\theta} g(\theta)\\
		&=
		-\frac{1}{n}\sum_{i=1}^n\int_0^\tau\left\{
		L_{i}-\bar{L}_n(t, \alpha, \beta)
		\right\}dN_i(t)+\lambda_{\theta} g(\theta),
	\end{split}
\end{align}
with $\lambda_{\theta}>0$ a penalty parameter \citep{Fu2003}, in Step 1 of the triple selection proposal. Here, $g(a)$ denotes a vector of elements $g(a_j)$, where $g(a_j) = |a_j|$ if $a_j\neq0$ and $g(a_j)\in [-1, 1]$ otherwise.
Likewise, we will solve the following penalized estimating equation for $\gamma$
\begin{align}\label{eq:subgrad_gamma}
\begin{split}
	0&=-\frac{1}{n}\sum_{i=1}^n\int_0^\tau\left[A_{i}-\bar{A}_n(t, \hat{\alpha}, \hat{\beta})-\gamma'\left\{
	L_{i}-\bar{L}_n(t, \hat{\alpha}, \hat{\beta})
	\right\} \right]\left\{
	L_{i}-\bar{L}_n(t, \hat{\alpha}, \hat{\beta})
	\right\}dN_i(t)\\
	&+\lambda_\gamma g(\gamma),
\end{split}
\end{align}
with $\lambda_{\gamma}>0$ a penalty parameter, in Step $3$ of the triple selection proposal. 
The penalty terms in Expression \eqref{eq:subgrad_beta} and \eqref{eq:subgrad_gamma} correspond to the subgradient of respectively the $l_1$ norm $||\theta||_1=||(\alpha, \beta)||_1$ with respect to $\theta=(\alpha, \beta)$ and the $l_1$ norm $||\gamma||_1$ with respect to $\gamma$; hence our procedure amounts to $l_1$-penalized $m$-estimation.
If $\lambda_{\theta}$ goes to zero as $n\to \infty$, 
Equation \eqref{eq:subgrad_beta} directly guarantees that $n^{-1}\sum_{i=1}^n\partial U_i(\alpha^*, \beta^*, \gamma^*)/\partial\gamma$ has approximately mean zero.
Denote the history spanned by $N(t)$ as $\mathcal{F}_t$. Then, if $\lambda_{\gamma}$ also goes to zero as $n\to \infty$, because the conditional mean $E\left\{dN(t)|A, L, \mathcal{F}_t\right\}$ equals $R(t)\lambda_0(t, \alpha^*, \beta^*)e^{\alpha^{*} A+\beta^{*'}L}$ under the assumption that the Cox model for survival time $T$ is correctly specified, we can show that Equation \eqref{eq:subgrad_gamma} renders the gradient $n^{-1}\sum_{i=1}^n\partial U_i(\alpha^*, \beta^*, \gamma^*)/\partial\beta$ approximately zero. We make a more rigorous argument in Appendix A of the online supplementary materials.
Under standard ultra-sparsity conditions (see Theorem \ref{th:main}) and the assumption that $\lambda_\theta=O\left(\sqrt{\log(p\lor n)/n}\right)$, $\lambda_\gamma=O\left(\sqrt{\log(p\lor n)/n}\right)$ and $\log(p\lor n)=o(n)$ (where $a\lor b$ denotes the maximum of $a$ and $b$), these properties help ensure that the second and third term in Equation \eqref{eq:taylor_alpha} are asymptotically negligible, regardless of the complex behavior of $\check{\beta}$ and $\check{\gamma}$. Our proposal is thus to select the variables in a specific way that shrinks the gradients in Equation \eqref{eq:taylor_alpha} to zero and consequently de-biases a na\"ive post-selection estimator (e.g., post-Lasso estimator).



\vspace{-0.5cm}
\begin{remark}[Model for censoring]\label{remark:cens}
	The discussion above suggests that Step 1 and Step 3 are all that is needed to make the gradients zero. 
	Specifically, fitting the exposure model as in Step $3$ of the triple selection approach makes the additional selection Step $2$ for censoring redundant.
	This is surprising and raises the question how a sufficient adjustment set for censoring can be guaranteed.
	To see this, consider the linear ``exposure'' model in Step $3$ which conditions on the risk set. As a consequence of collider-stratification, predictors of censoring may become dependent on the exposure in the risk set.
		\begin{figure}[h!]
		\centering
		\includegraphics[width=0.5\textwidth]{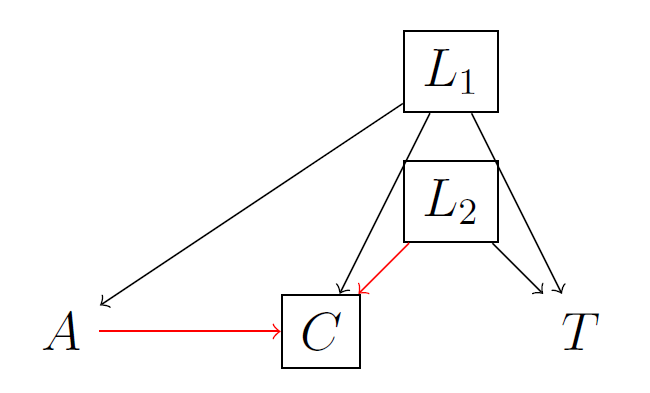}
		\caption{\label{fig:DAG} A Directed Acyclic Graph (DAG) to explain why exposure model also picks up predictors of censoring.}
	\end{figure}
	Indeed, conditioning on the risk set in particular implies conditioning on being uncensored. From the  Directed Acyclic Graph (DAG) in Figure \ref{fig:DAG}, it is clear that conditioning on $C$ opens a path between $A$ and $L_2$, making them associated in the absence of a direct effect of $L_2$ on $A$. One may therefore detect predictors of exposure and censoring via one model for exposure in the risk set. 
	Even so, since collider-stratification usually only induces weak associations \citep{Greenland2003}, we expect better finite sample performance when an additional (direct) model for censoring is used.
	This has the further advantage that it considerably simplifies the procedure in settings where exposure is independent of baseline covariates as Step 3 is then superfluous. This is in particular the case in randomized experiments \citep[see][]{vanlancker2020}.
\end{remark}

\begin{remark}[Link to poor man's approach]
	Although the model for censoring (Step 2) in the triple selection approach is in principle redundant, it is essential in the poor man's approach to control for informative censoring as the exposure model is no longer conditional on the risk set.
	By fitting a potentially correct model for the exposure assignment mechanism at baseline, we expect the poor man's approach to pick up more variables than the triple selection approach. 
	We therefore view it as an under-smoothed version of the triple selection approach, which may justify its validity.
\end{remark}


\subsubsection{Asymptotic Properties}
We will now study convergence of $\check{\alpha}$ more formally under a sequence of laws $P_n\in \mathcal{P}$,
where $\mathcal{P}$ represents a class of observed data laws that obey a correctly specified Cox proportional hazards model for survival time $T$ conditional on $A$ and $L$. 
We will allow for $p$ to increase with $n$, and for the values of the parameters $\gamma$, $\beta$ and $\eta_2$ to depend on $n$. This is done in
order to gain better insight into the finite-sample behavior of the test statistic when the data-generating mechanism changes with $n$ (e.g., $p$ grows with $n$). 
Let $\mathbb{P}_{P_{n}}$ and $\mathbb{E}_{P_{n}}$ respectively denote a probability and expectation taken with respect to this local data-generating process $P_n$.
Let $\alpha_n$ and $\beta_n$ denote the population values of $\check{\alpha}$ and $\check{\beta}$, respectively. 
Moreover, define 
\[\bar{A}^*(t, \alpha, \beta)=\frac{\mathbb{E}_{P_n}\left\{AR(t) e^{\alpha A+\beta'L}\right\}}{\mathbb{E}_{P_n}\left\{R(t) e^{\alpha A+\beta'L}\right\}}\quad\textrm{and}\quad \bar{L}^*(t, \alpha, \beta)=\frac{\mathbb{E}_{P_n}\left\{LR(t) e^{\alpha A+\beta'L}\right\}}{\mathbb{E}_{P_n}\left(R(t) e^{\alpha A+\beta'L}\right)}.\]
The population value $\gamma_n$ of $\check{\gamma}$ is then defined as the population coefficient from a least squares regression of $A_i-\bar{A}^*(T_i, \alpha_n, \beta_n)$ on $L_i-\bar{L}^*(T_i, \alpha_n, \beta_n)$ in subjects for whom an event was observed during the study.  
Define $\check{\eta}_2$ as the estimate of $\eta_2$ obtained by fitting a Cox model for censoring time $C$ given exposure $A$ and the union of variables estimated to have non-zero coefficients in the first three steps of the triple selection approach. The population value of $\check{\eta}_2$ is denoted as $\eta_{2n}$.


\vspace{-0.5cm}
\begin{theorem}[Robust Estimation and Inference]\label{th:main}	
	Define the sets of variables that have coefficients that are truly non-zero (i.e., active set) as $S_\gamma = \text{support}(\gamma_n)$, $S_\beta = \text{support}(\beta_n)$ and $S_\eta = \text{support}(\eta_{2n})$. Moreover, let $p$ denote the number of covariates, and define $s_\gamma=|S_\gamma|$, $s_\beta=|S_\beta|$ and $s_\eta=|S_\eta|$. Suppose the ultra-sparsity condition $n^{-1/2}(s_\gamma\lor s_\beta\lor s_\eta)\log(p\lor n)=o(1)$ holds. 
	Then, under Assumptions 1-8 in the online supplementary materials and the primitive conditions required for uniform consistency of the Lasso estimators (see e.g., Assumptions $2$ and $5$ in \citet{fang2017}), including
	$\lambda_\theta=O\left(\sqrt{\frac{\log(p\lor n)}{n}}\right)$ and $\lambda_\gamma=O\left(\sqrt{\frac{\log(p\lor n)}{n}}\right)$,
	the post-triple selection estimator $\check{\alpha}$ obeys 
	\begin{align}
		\sqrt{n}\left(\check{\alpha}-\alpha_n\right)=Z_{n}+o_{P_n}(1), \quad Z_{n} \stackrel{d}{\rightarrow} N(0,\Sigma_{n}^2)\label{eq:theorem1}
	\end{align}
	as $n \rightarrow \infty$. Here, $o_{P_n}(1)$ denotes a stochastic variable that converges in probability to zero as $n \rightarrow \infty$ under the measure $P_n$, and
	$$
	Z_{n}:=\frac{V^{-1}}{\sqrt{n}} \sum_{i=1}^{n}\int_0^\tau\left[A_{i}-\bar{A}^*(t, \alpha_n, \beta_n)-\gamma_n'\left\{
	L_{i}-\bar{L}^*(t, \alpha_n, \beta_n)
	\right\} \right]dM_i(t),
	$$
	$$ \Sigma_{n}^{2}:=V^{-1}\mathbb{E}_{P_n}\left\{\left(\int_0^\tau\left[A_{i}-\bar{A}^*(t, \alpha_n, \beta_n)-\gamma_n'\left\{
	L_{i}-\bar{L}^*(t, \alpha_n, \beta_n)
	\right\} \right]dM_i(t)\right)^2\right\}V^{-1},
	$$
	and
	\begin{align*}
		V:=&\mathbb{E}_{P_n}\left(\int_{0}^{\tau}\left[A_{i}-\bar{A}^*(t, \alpha_n, \beta_n)-\gamma_n'\left\{
		L_{i}-\bar{L}^*(t, \alpha_n, \beta_n)
		\right\} \right]R_i(t)e^{\alpha_n A_i+\beta_n' L_i}\lambda_{0n}(t, \alpha_n, \beta_n)\right.\\
		&\hspace{1cm}\left.\left\{A_i-\bar{A}^*(t, \alpha_n, \beta_n) \right\}dt\vphantom{\int_{0}^{\tau}}\right),
	\end{align*}
	where $$dM_i(t)=dN_i(t)-\lambda_{0n}(t, \alpha_n, \beta_n)e^{\alpha_nA_i+\beta_n' L_i}R_i(t)dt$$ and $$\lambda_{0n}(t, \alpha, \beta)=\mathbb{E}_{P_n}\{R(t)dN(t)\}\mathbb{E}_{P_n}^{-1}\{R(t)e^{\alpha A+\beta' L}\}.$$
	
	Thus, $\Sigma_{n}^{-1}\sqrt{n}\left(\check{\alpha}-\alpha_n\right)$ converges weakly to $N(0,1)$. 
	Moreover, $\Sigma_{n}^{2}$ can be consistently estimated as
		\begin{align*}
		\widehat{\Sigma}_{ n}^{2}=&\widehat{V}^{-1}\mathbb{E}_{n}\left\{\left(\int_0^\tau\left[A_{i}-\bar{A}_n(t, \check{\alpha}, \check{\beta})-\check{\gamma}'\left\{
		L_{i}-\bar{L}_n(t, \check{\alpha}, \check{\beta})
		\right\} \right]dM_i(t)\vphantom{\int_{0}^{\tau}}\right)^2\right\}\widehat{V}^{-1},
	\end{align*}
	with 
	\begin{align*}
		\widehat{V}=&\mathbb{E}_{n}\left(\int_0^\tau\left[A_{i}-\bar{A}_n(t, \check{\alpha}, \check{\beta})-\check{\gamma}'\left\{
		L_{i}-\bar{L}_n(t, \check{\alpha}, \check{\beta})
		\right\} \right]R_i(t)e^{\check{\alpha} A_i+\check{\beta}' L_i}\hat{\lambda}_0(t, \check{\alpha}, \check{\beta})\right.\\
		&\hspace{1cm}\left.\left\{A_{i}-\bar{A}_n(t, \check{\alpha}, \check{\beta}) \right\}dt\vphantom{\int_{0}^{\tau}}\right).
	\end{align*}
\end{theorem}

The key ultra-sparsity condition $n^{-1/2}(s_\gamma\lor s_\beta\lor s_\eta)\log(p\lor n)=o(1)$ demands the number of non-zero coefficients in the models for $\gamma$, $\beta$ and $\eta$ to be small relative to the square
root of the overall sample size. Such assumption is common in the literature on high-dimensional inference \citep{belloni2016, ning2017}.
The theorem allows for the data-generating processes to change with $n$, in particular allowing sequences
of regression models with coefficients never perfectly distinguishable from zero, i.e., models where perfect
model selection is not possible. In doing so, the results achieved in Theorem \ref{th:main} imply validity
uniformly over a large class of sparse models, which is referred to as ``honesty'' in the statistical literature \citep[see e.g.,][]{Li1989honest}. In particular, as stated in the following corollary, our
proposed test and confidence intervals are uniformly valid over the parameter space under the ultra-sparsity condition.
The addition of variables that are purely predictive of censoring may impact the asymptotic behavior of the estimator when for instance $A$ has no effect on censoring, since these variables then no longer reduce bias but only increase the variance of the estimator. This will nevertheless be accounted for in the asymptotic distribution of the estimator, since the Schoenfeld residual terms in the decorrelated score equation will have reduced variability. Otherwise, the impact in large samples should be negligible since we assume sparsity in the censoring model. Hence, we do not explictly account for Step 2 of the triple selection procedure in the proofs.

\begin{Corollary}[Uniform $\sqrt{n}$-Rate of Consistency and Uniform Normality]
	Under Assumptions 1-8 in the online supplementary materials, the primitive conditions required for uniform consistency of the Lasso estimators and the ultra-sparsity condition $n^{-1/2}(s_\gamma\lor s_\beta\lor s_\eta)\log(p\lor n)=o(1)$, the post-triple selection estimator, $\check{\alpha}$, is $\sqrt{n}$-consistent and asymptotically normal
	uniformly over $\mathcal{P},$ namely
	$
	\lim _{n \rightarrow \infty} \sup _{P_{n} \in \mathcal{P}} \sup _{r \in \mathbb{R}}\left|\mathbb{P}_{P_{n}}\left(\Sigma_{n}^{-1} \sqrt{n}\left(\check{\alpha}-\alpha_n\right) \leqslant r\right)-\Phi(r)\right|=0.
	$
	Moreover, the result continues to hold if $\Sigma_{n}$
	is replaced by $\hat{\Sigma}_{n}$ specified in the statement of the previous theorem. 
\end{Corollary} 
This result can be used to show the uniform validity of the proposed tests and confidence intervals, as in \cite{belloni2014}.

\section{Simulation Study}\label{sec:simulation}
In this section, we examine the finite-sample properties of the poor man's approach and the post-triple selection test. In particular, we wish to compare their performance to that of the method proposed in \citet{fang2017} and a standard post-single selection method (i.e., post-Lasso). The latter refits the Cox model to the variables selected by the first-step penalized variable selection method (i.e., Lasso; this method omits Step 2 and Step 3 of the strategies proposed in this paper). We present the simulation study as recommended in \cite{Morris2019}.

\subsection{Simulation Design}

\hspace{0.3cm}\textbf{Aims}:
To evaluate and compare the finite-sample Type I error rate of the test statistic based on the proposed poor man's approach, the proposed triple selection strategy, the post-Lasso approach and the decorrelated score function in \citet{fang2017}.

\textbf{Data-Generating Mechanisms}:
In each setting, we generate $n$ mutually independent vectors $\left(T_{i}, C_{i}, A_{i}, L_{i}'\right)', i=1, \ldots, n .$ Here, we consider the following data-generating mechanisms for $(A_i, L_i')'=X_i$:
\begin{enumerate}[(a)]
	\item $X_i\sim N_{p+1}\left(\mathbf{0}, \Sigma\right)$, where $\Sigma$ is a Toeplitz matrix with $\Sigma_{jk}=\rho^{\left|j-k\right|}$ and $\rho=0.25$ and $0.50$,
	\item $L_i\sim N_p(\mathbf{0}, \mathbb{I})$ and
	$A_i\sim N(\nu_AL, 1), \text{ with } \nu_A=c_A(1, 1/2, \dots, 1/10, 0_{11}, \dots, 0_{p})'$,
\end{enumerate}
with $c_A$ a scalar parameter.
The $i$th survival and censoring time are based on respectively $T_i\sim\exp(\lambda_{T,i})$ with $\lambda_{T,i}=\exp(\beta_0)\exp(\alpha A_i+\beta' L_i)$ and $C_i\sim\exp(\lambda_{C,i})$ with $\lambda_{C,i}=\exp(\eta_0)\exp(\eta_1 A_i+\eta_2'L_i)$,
where $\beta_0$, $\alpha$, $\eta_0$ and $\eta_1$ are scalar parameters, and $\beta=b\cdot\nu_T$ and $\eta_2=g\cdot\nu_C$ are $p$-dimensional parameters with $b$ and $g$ scalar and $\nu_T$ and $\nu_C$ $p$-dimensional parameters. In our simulation study, we consider the following coefficient vectors $\nu_T$ and $\nu_C$
\begin{enumerate}
\vspace{-0.25cm}
	\item $\nu_T=(1, 1/2, \dots,1/9, 1/10, 0_{11}, \dots, 0_p)',$\\
	$\nu_C=(1, 1/2, 1/3, 1/4, 1/5, 1, 1/2,1/3, 1/4, 1/5, 0_{11}, \dots, 0_p)',$
	\item $\nu_T=(1, 1/2, \dots,1/9, 1/10, 0_{11}, \dots, 0_p)',$\\
	$\nu_C=(1, 1/2, 1/3, 1/4, 1/5, 0_6, \dots, 0_{10}, 1, 1/2,1/3, 1/4, 1/5, 0_{16}, \dots, 0_p)',$\vspace{-0.25cm}
\end{enumerate}
where the subscripts indicate the index (i.e., position) of $0$ in the vector. 
The parameters $b$ and $g$ are used to control the association between the covariates and respectively the survival and censoring times. Increments of $0.25$, starting from $0$ to $2$, are considered.
The coefficients $\beta_0$ and $\eta_0$ are set to $0$, corresponding with a baseline hazard of 1.

As we are interested in the Type I error of a test of the null hypothesis that $\alpha=0$ versus the alternative hypothesis that $\alpha\neq 0$ based on the different methods, $\alpha$ is set to $0$. The level of significance is set to be $0.05$.
For the data generating mechanisms described
above, we perform $1,000$ Monte Carlo runs for $n = 400$ and $p = 30$. As an objective choice we have chosen $\tau=+\infty$.

\textbf{Targets:} Our target of interest is the null hypothesis of no effect of treatment $A$ on the considered survival endpoint at a $5\%$ significance level.

\textbf{Methods of Analysis:}
Each simulated dataset is analyzed using the following methods:
\begin{itemize}
	\item Triple selection: Wald test based on robust SE for the estimator for $\alpha$ obtained via the post-triple selection approach proposed in Section \ref{sec:debiased}.
	\item Poor man's approach: Wald test based on robust SE for the estimator for $\alpha$ obtained via the poor man's approach proposed in Section \ref{sec:poor}.
	\item Post-Lasso: Wald test based on robust SE for the estimator for $\alpha$ obtained via post-Lasso, which refits the Cox model to the variables selected  in a Cox model for the outcome with Lasso (i.e., this method omits Step 2 and Step 3 of the strategies proposed in this paper). 
	\item Decorrelated test in \citet{fang2017}: test based on the estimator for $\alpha$ obtained via the the decorrelated score function in \citet{fang2017} using Lasso. \vspace{-0.25cm}
\end{itemize}

All considered approaches require the selection of penalty parameters. 
In this simulation study, we use a $20$-fold cross-validation technique with the negative cross-validated penalized (partial) log-likelihood as loss function. We obtain the penalty parameter $\lambda_\text{1se}$, which is the largest value of the penalty parameter such that the cross-validated error is within $1$ standard error of the minimum, using the function \texttt{cv.glmnet()} in the \texttt{R} package \texttt{glmnet}.

\textbf{Performance Measures:} We assess the finite-sample (empirical) Type I error rate of the test of no exposure effect on the considered survival endpoint $T$. 

\subsection{Simulation Results}

\begin{figure}[h!]
	\begin{subfigure}{.45\linewidth}
		\centering
		\includegraphics[width=\textwidth]{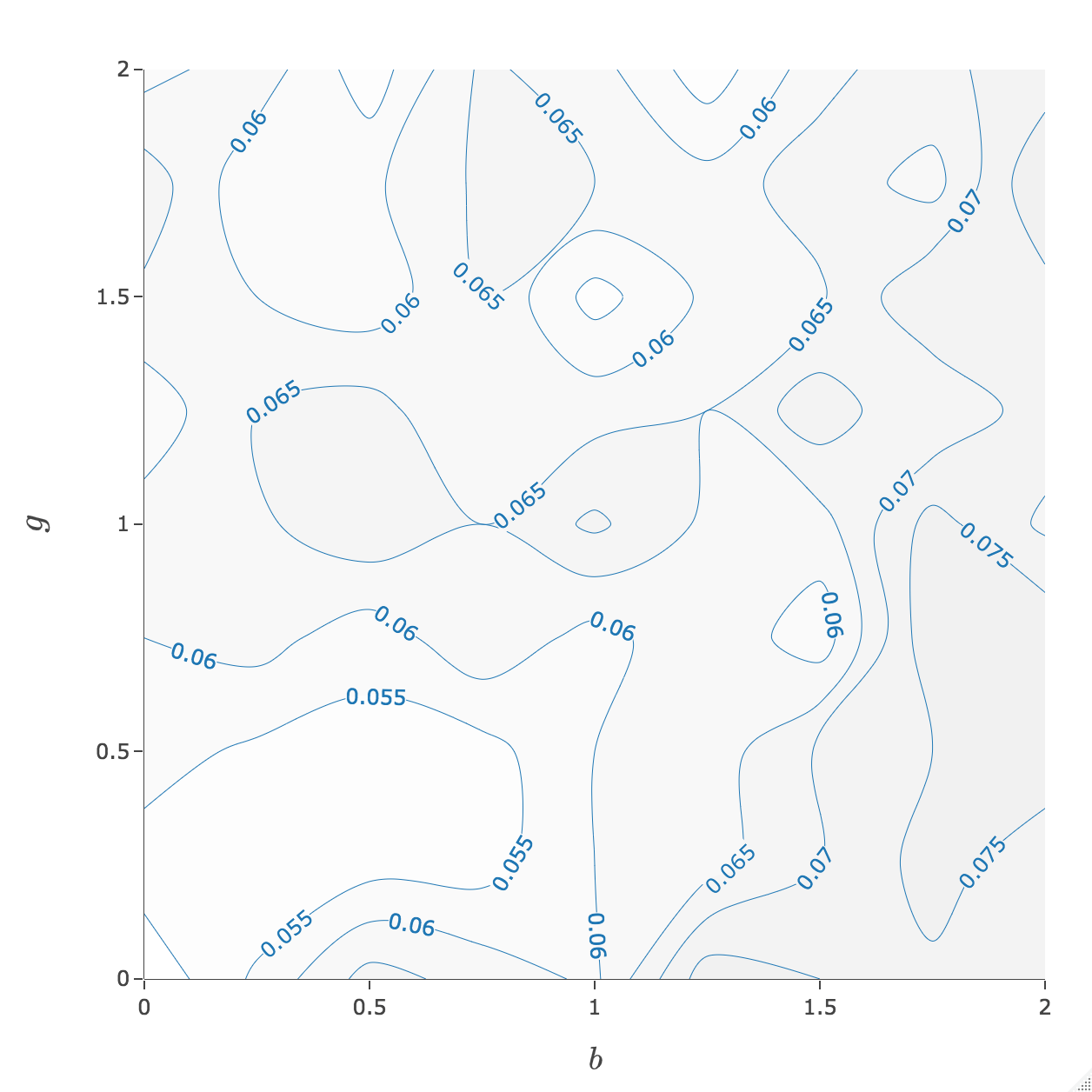}
		\caption{Triple selection} \label{fig:Setting1_1_Prop}
	\end{subfigure}\hfill
	\begin{subfigure}{.45\linewidth}
		\centering
		\includegraphics[width=\textwidth]{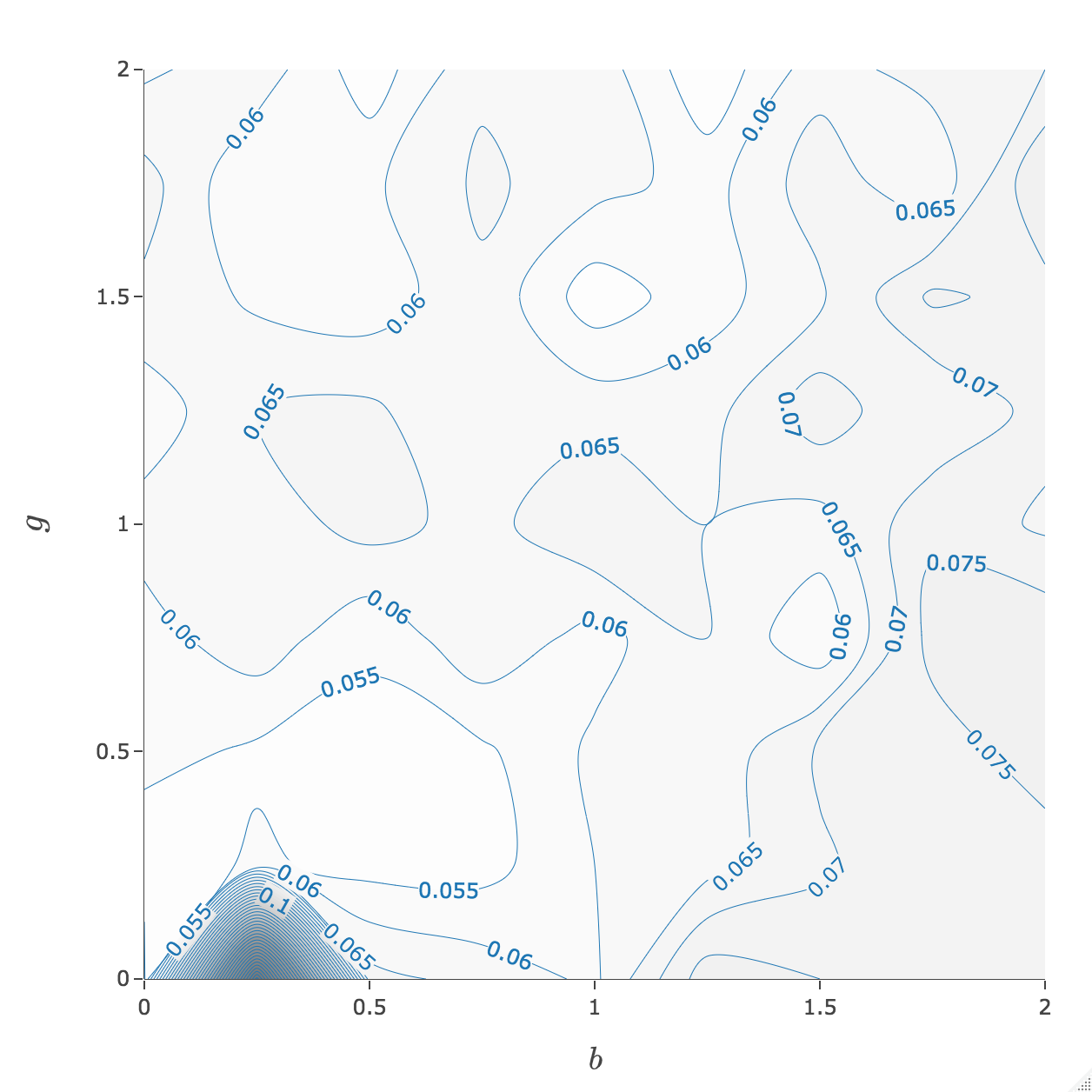}
		\caption{Poor man's approach} \label{fig:Setting1_1_PMA}
	\end{subfigure}\hfill	
	\begin{subfigure}{.45\linewidth}
		\centering
		\includegraphics[width=\textwidth]{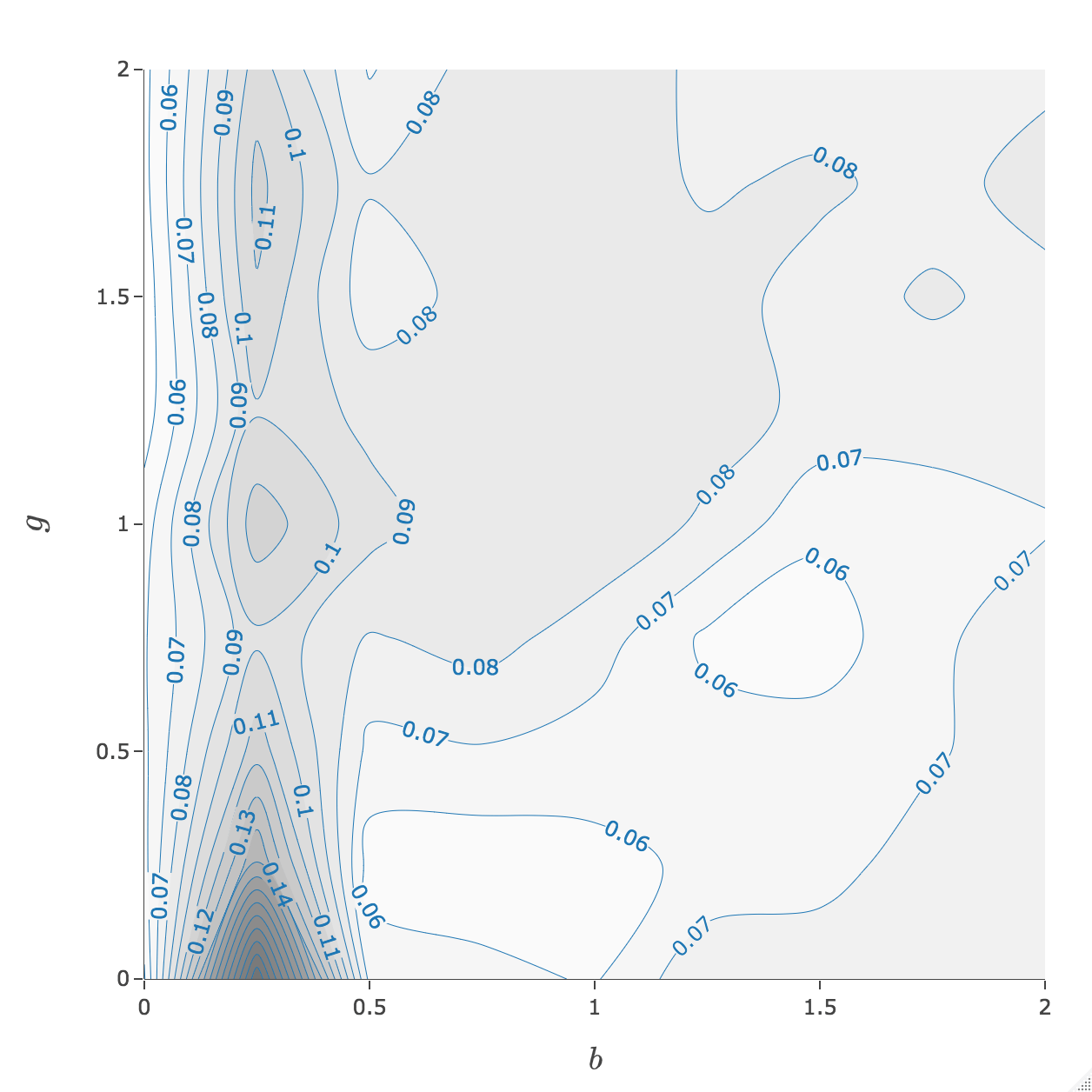}
		\caption{Post-Lasso} \label{fig:Setting1_1_Lasso}
	\end{subfigure}\hfill
	\begin{subfigure}{.45\linewidth}
		\centering
		\includegraphics[width=\textwidth]{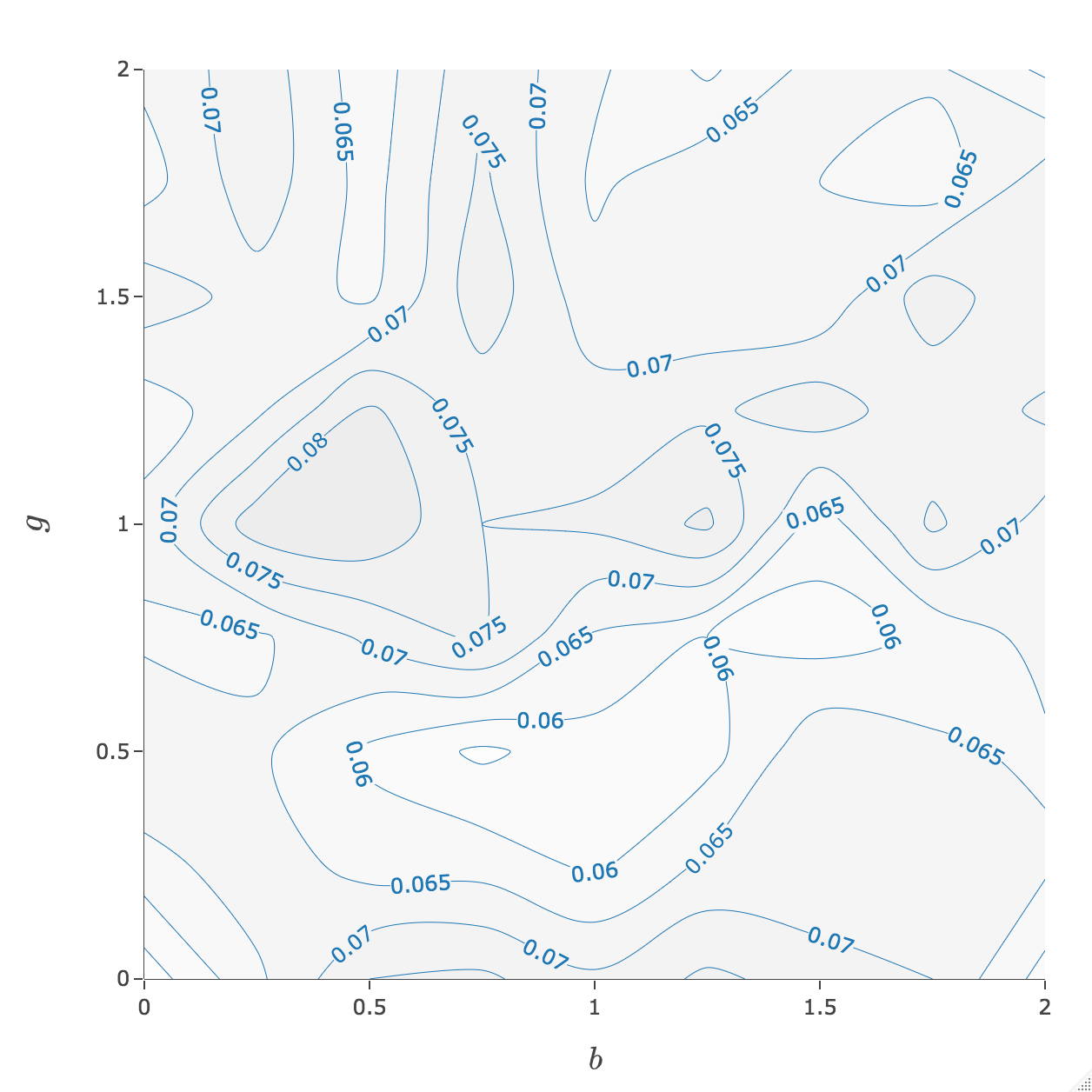}
		\caption{Decorrelated test in \citet{fang2017}} \label{fig:Setting1_1_Fang}
	\end{subfigure}\hfill
	\caption{Empirical Type I error rate at the $5\%$ significance level of the different tests under Setting 1(a) with $n=400$, $p=30$, $\rho=0.50$, $\eta_1=1$, $\beta_0=0$ and $\eta_0=0$. 
	} \label{fig:Setting1_1}
\end{figure}

\begin{figure}[h!]
	\begin{subfigure}{.45\linewidth}
		\centering
		\includegraphics[width=\textwidth]{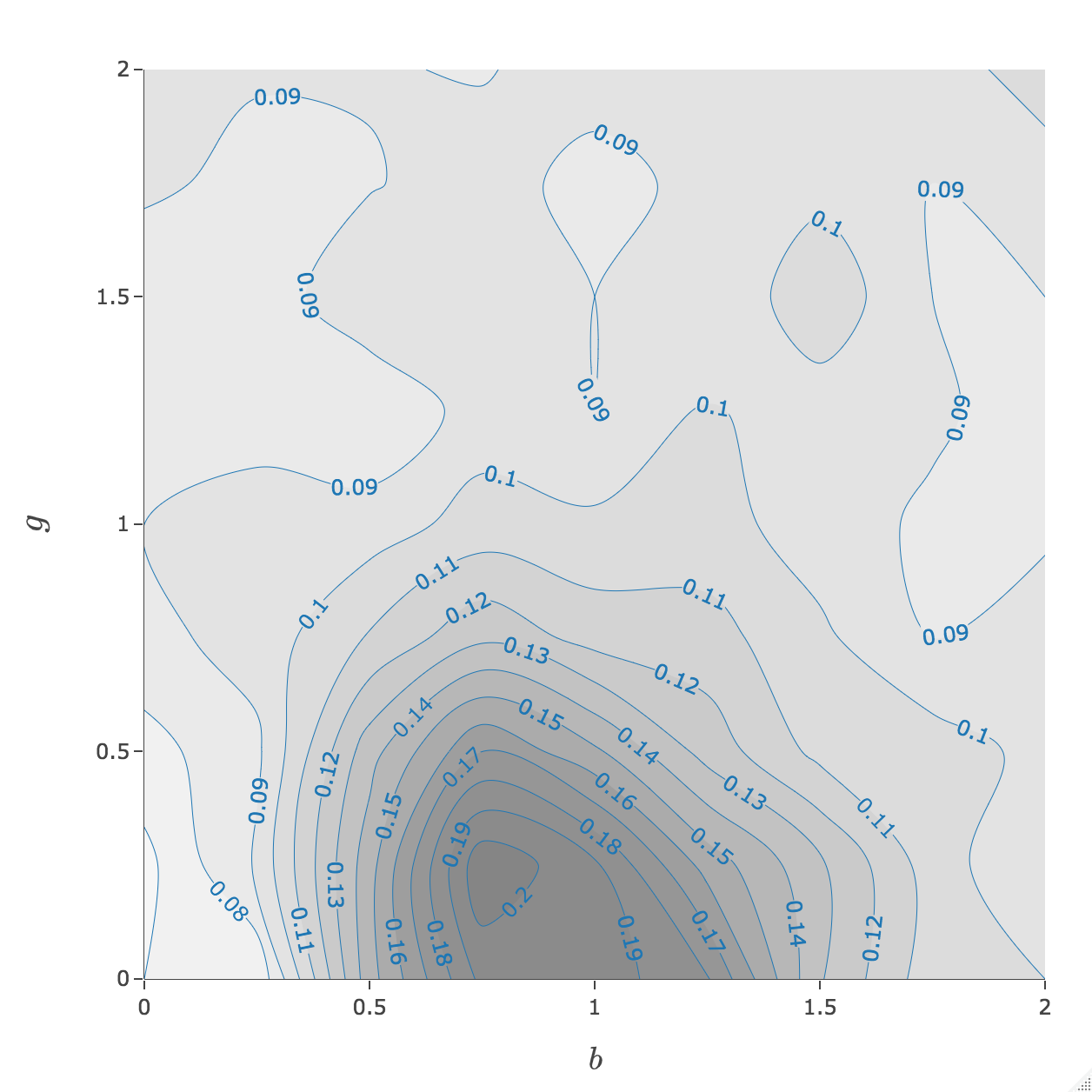}
		\caption{Triple selection} \label{fig:Setting1_1d_Prop}
	\end{subfigure}\hfill
	\begin{subfigure}{.45\linewidth}
		\centering
		\includegraphics[width=\textwidth]{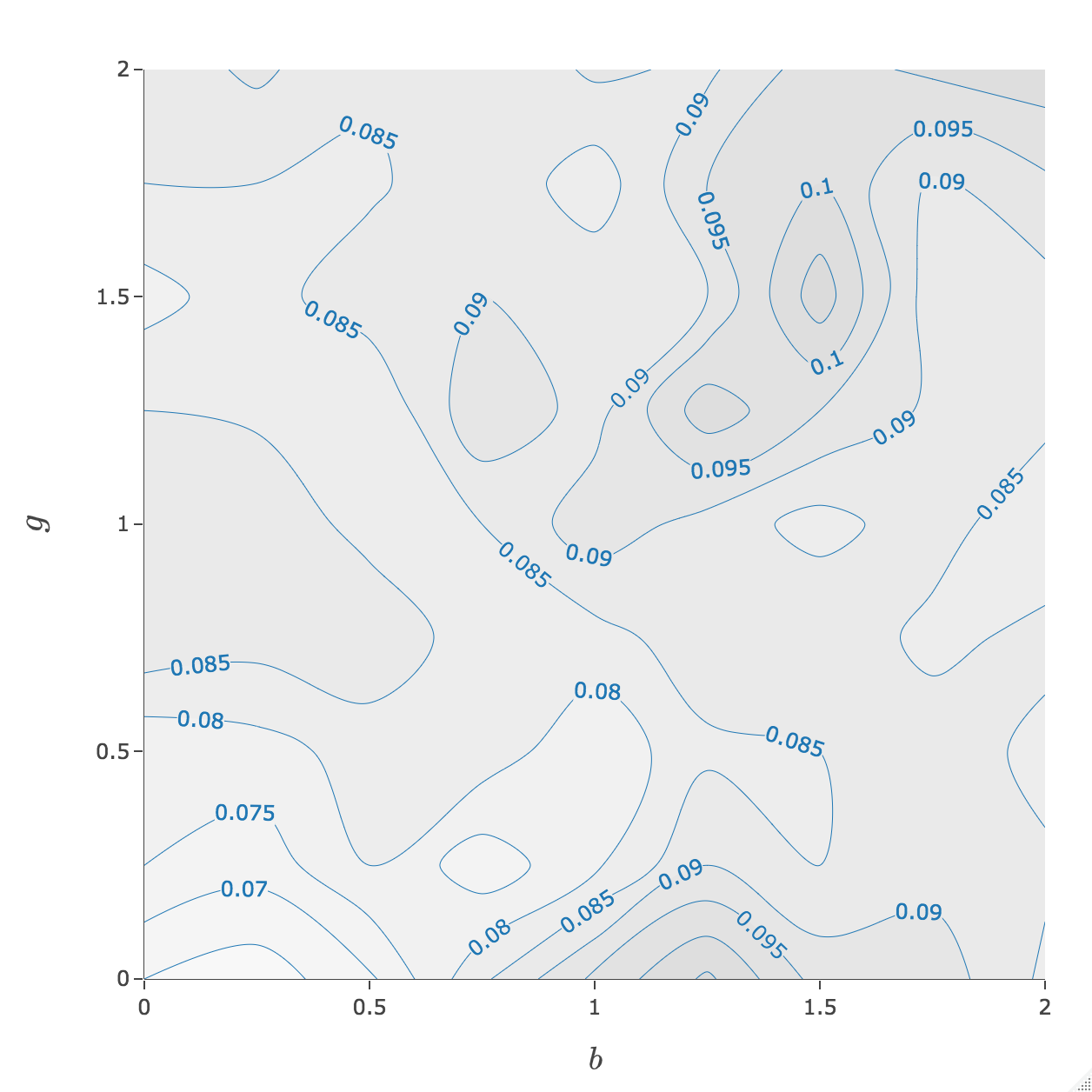}
		\caption{Poor man's approach} \label{fig:Setting1_1d_PMA}
	\end{subfigure}\hfill	
	\begin{subfigure}{.45\linewidth}
		\centering
		\includegraphics[width=\textwidth]{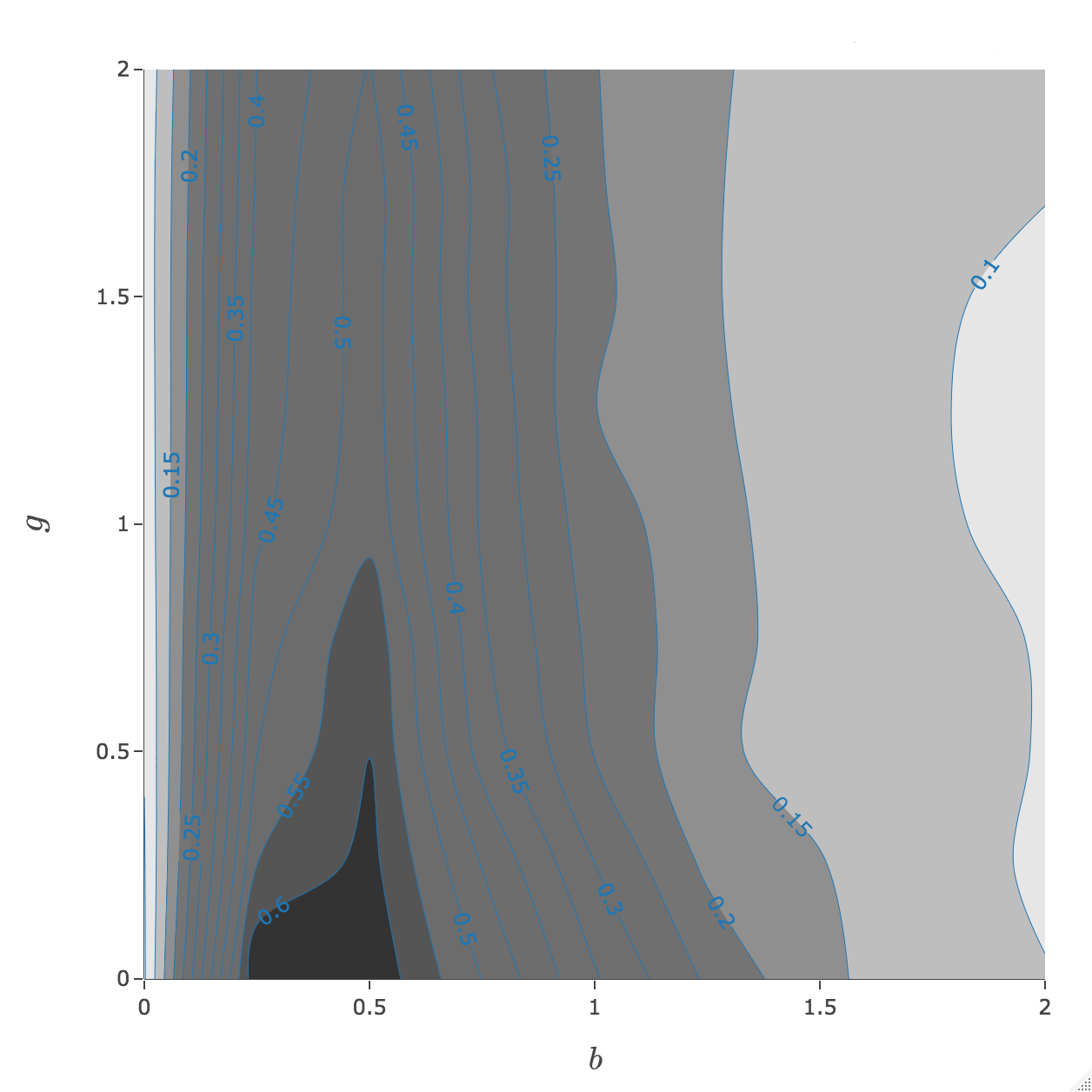}
		\caption{Post-Lasso} \label{fig:Setting1_1d_Lasso}
	\end{subfigure}\hfill
	\begin{subfigure}{.45\linewidth}
		\centering
		\includegraphics[width=\textwidth]{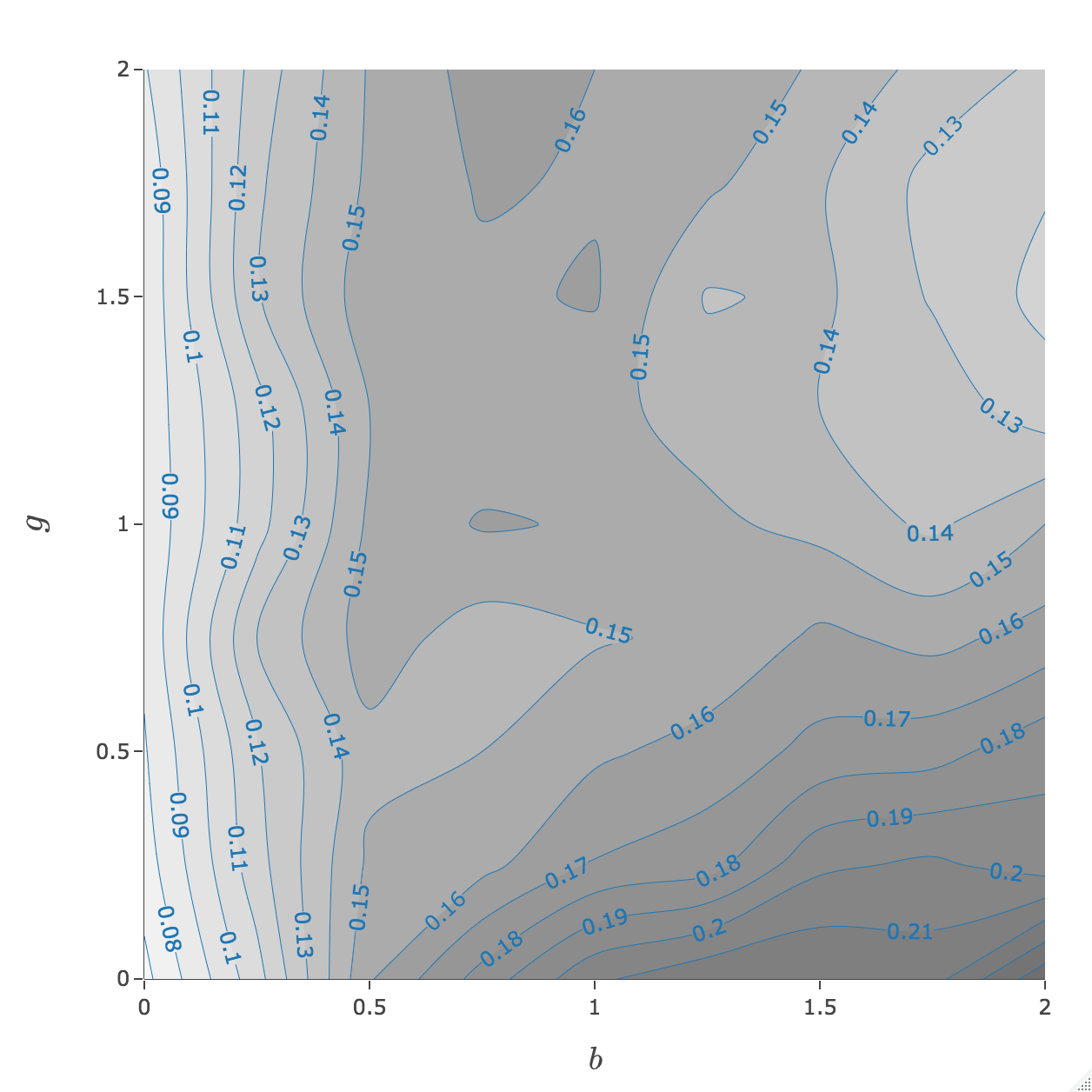}
		\caption{Decorrelated test in \citet{fang2017}} \label{fig:Setting1_1d_Fang}
	\end{subfigure}\hfill
	\caption{Empirical Type I error rate at the $5\%$ significance level of the different tests under Setting 1(b) with $n=400$, $p=30$, $c_A=1$, $\eta_1=1$, $\beta_0=0$ and $\eta_0=0$. 
	} \label{fig:Setting1_1d}
\end{figure}

\begin{figure}[h!]
	\begin{subfigure}{.45\linewidth}
		\centering
		\includegraphics[width=\textwidth]{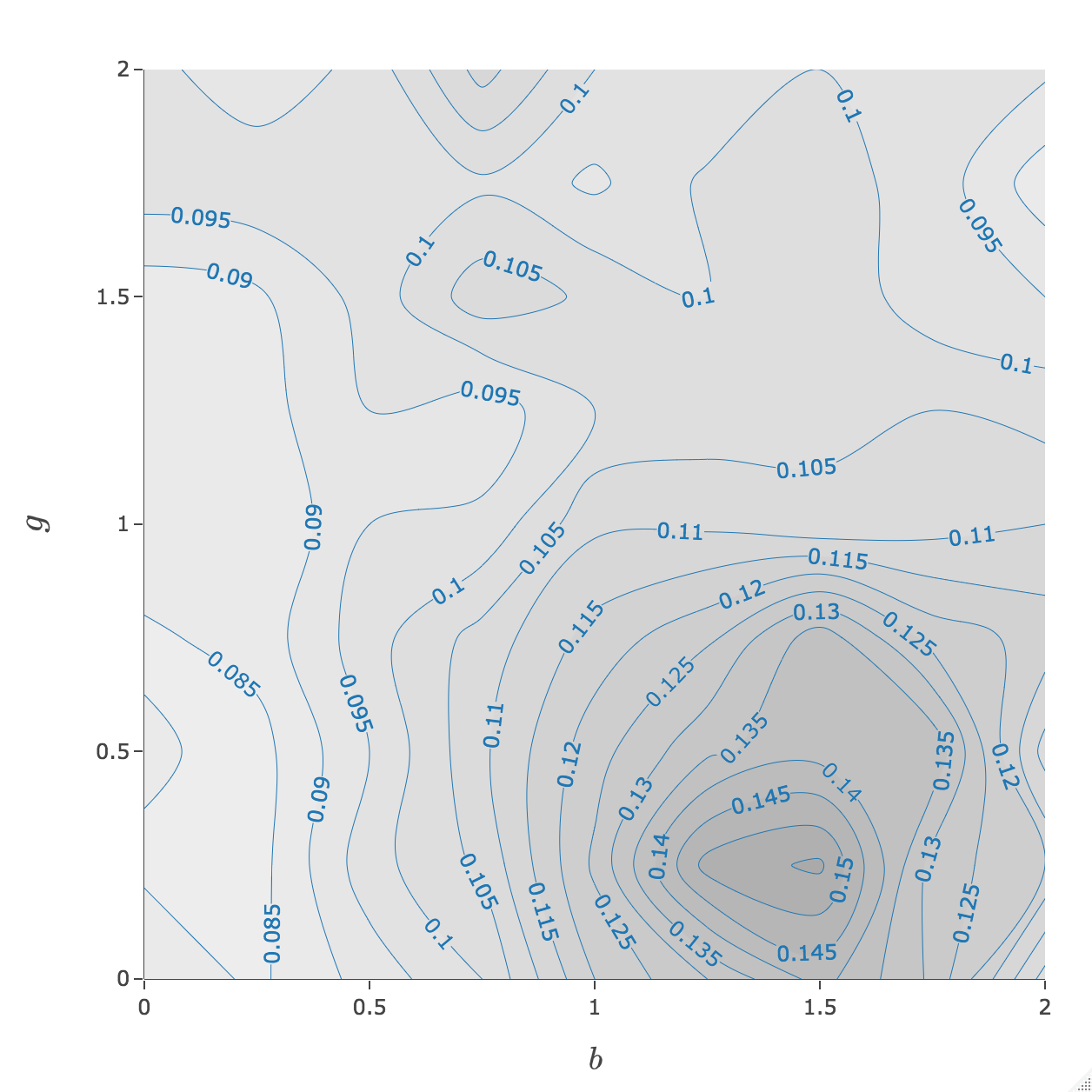}
		\caption{Triple selection} \label{fig:Setting1_1e_Prop}
	\end{subfigure}\hfill
	\begin{subfigure}{.45\linewidth}
		\centering
		\includegraphics[width=\textwidth]{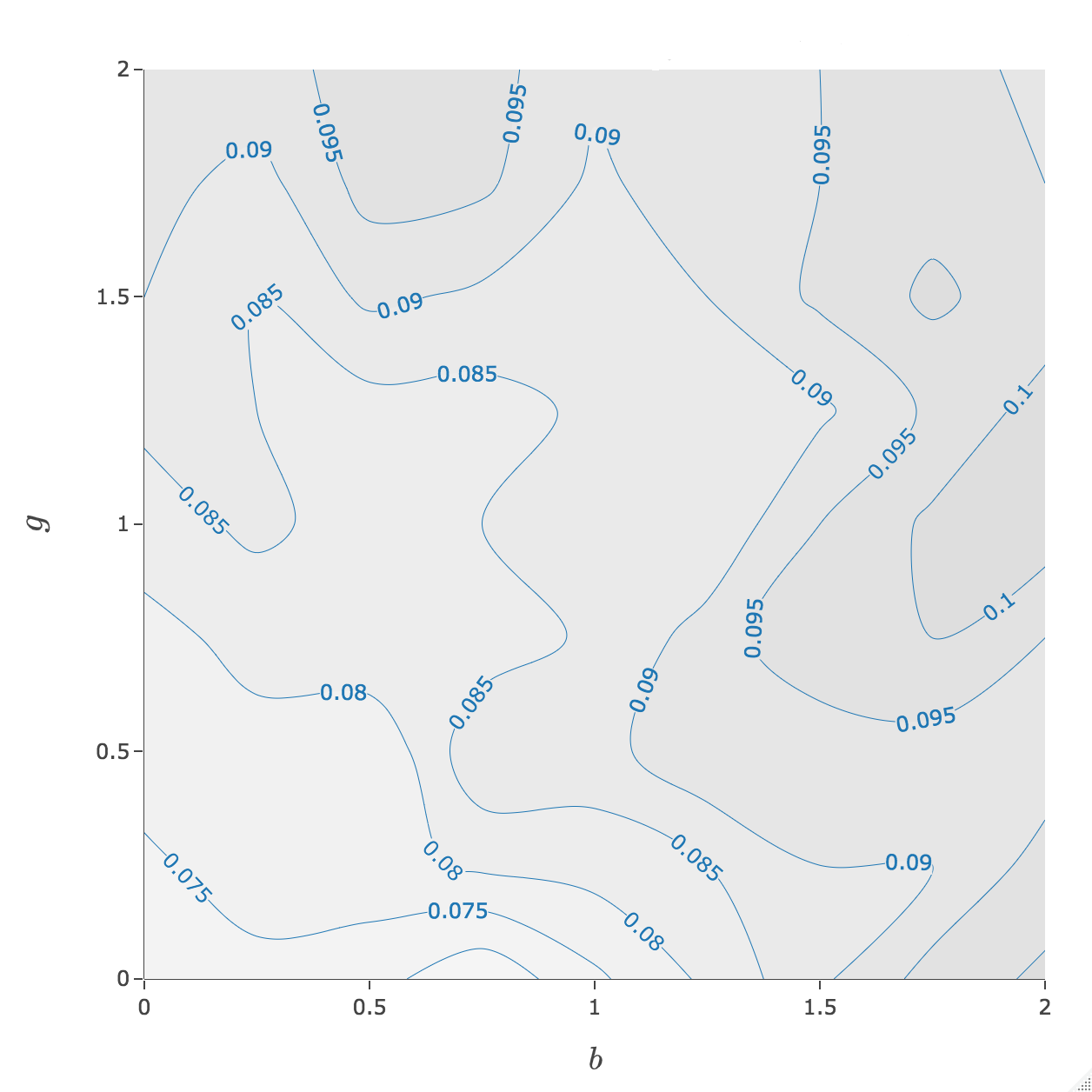}
		\caption{Poor man's approach} \label{fig:Setting1_1e_PMA}
	\end{subfigure}\hfill	
	\begin{subfigure}{.45\linewidth}
		\centering
		\includegraphics[width=\textwidth]{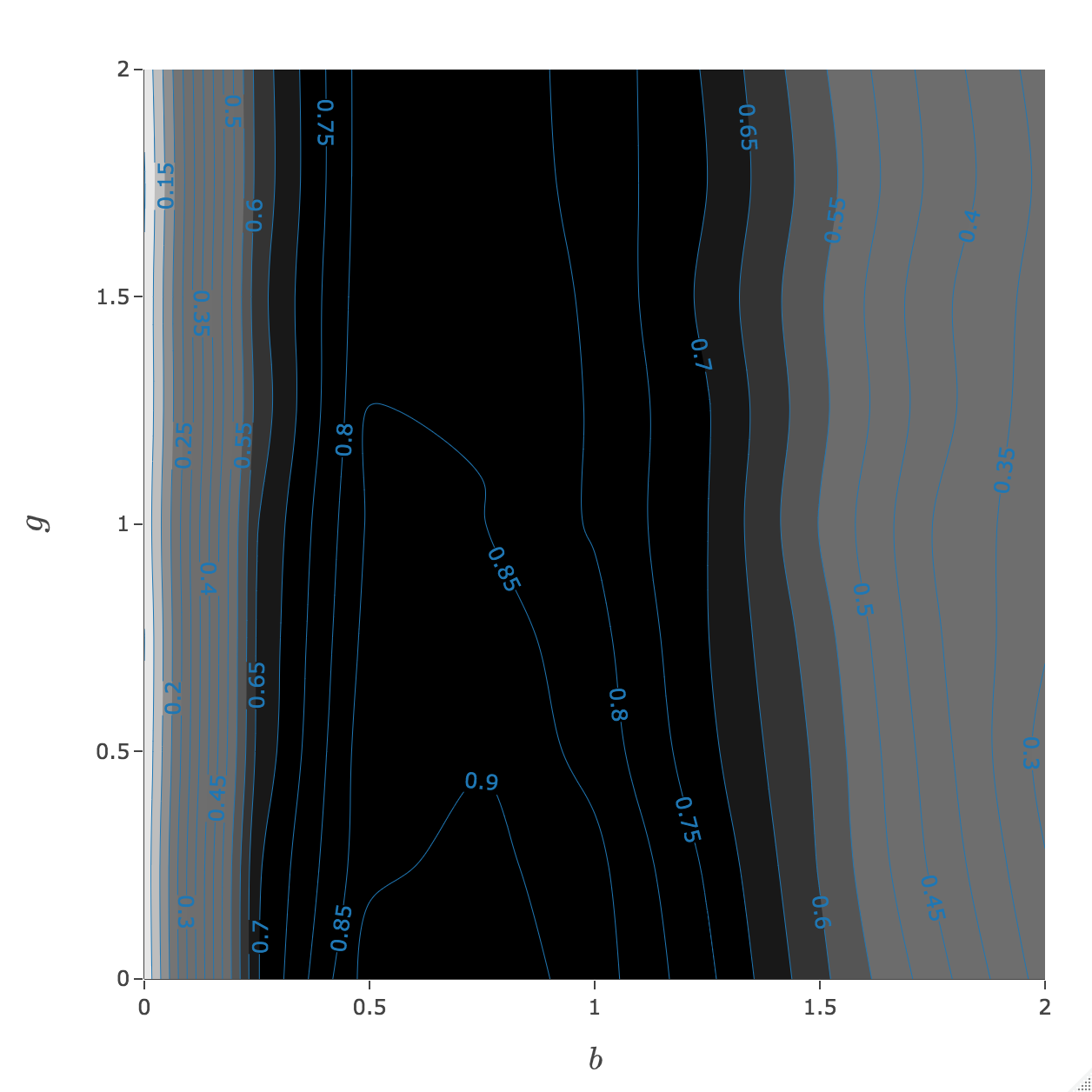}
		\caption{Post-Lasso} \label{fig:Setting1_1e_Lasso}
	\end{subfigure}\hfill
	\begin{subfigure}{.45\linewidth}
		\centering
		\includegraphics[width=\textwidth]{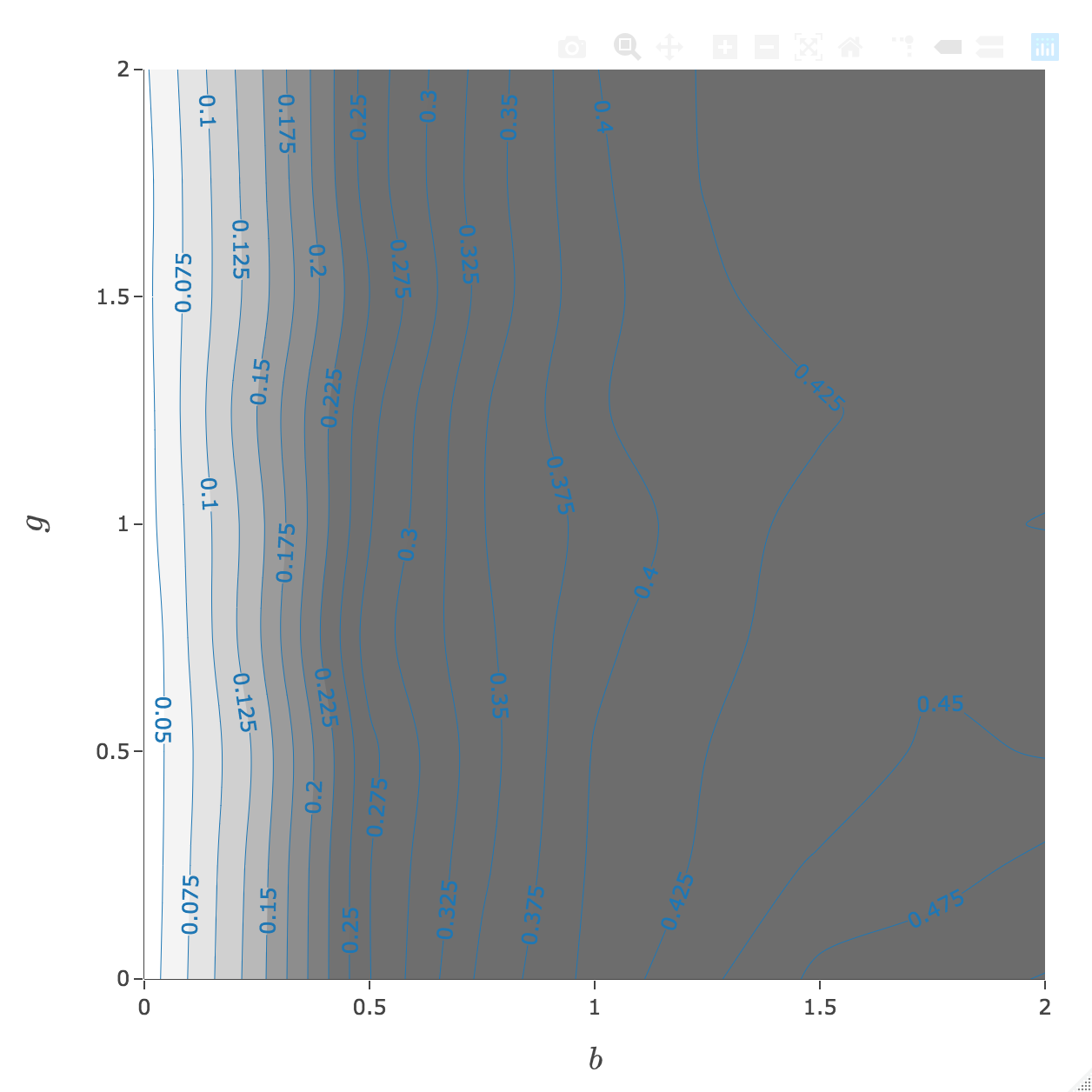}
		\caption{Decorrelated test in \citet{fang2017}} \label{fig:Setting1_1e_Fang}
	\end{subfigure}\hfill
	\caption{Empirical Type I error rate at the $5\%$ significance level of the different tests under Setting 1(b) with $n=400$, $p=30$, $c_A=2$, $\eta_1=1$, $\beta_0=0$ and $\eta_0=0$. 
	} \label{fig:Setting1_1e}
\end{figure}

The empirical Type I errors for the 
different tests
under Setting $1$ with $\eta_1=1$ are summarized in Figure \ref{fig:Setting1_1} for Setting (a) with $\rho=0.50$, Figure \ref{fig:Setting1_1d} for Setting (b) with $c_A=1$ and Figure \ref{fig:Setting1_1e} for Setting (b) with $c_A=2$. 

The simulation results show that over the different settings, the poor man's approach has a rejection rate closest to the nominal level of $5\%$. 
In sharp contrast, the Type I errors based on the post-Lasso selection perform very poorly, and deviate strongly away from the nominal level of $0.05$ throughout large parts of the model space (see Figures \ref{fig:Setting1_1d} and \ref{fig:Setting1_1e}). This is a result of eliminating too many covariates from the outcome model. Although the proposal by \citet{fang2017} clearly provides lower rejection rates than the post-Lasso approach, it is outperformed by both proposals throughout large parts of the model space (see Figures \ref{fig:Setting1_1d} and \ref{fig:Setting1_1e}). This might occur because re-fitting the Cox model, adjusting for the union of covariates selected in three steps (as is done in the triple selection and poor man's approaches), may take the empirical analogues of the gradients from the Taylor expansion in Section \ref{sec:intuition} closer to zero within the sample. Similar results were also observed in \cite{belloni2016}. This is moreover in line with our expectation of better finite sample performance when an additional model for censoring is used (see Remark \ref{remark:cens}).

The Type I errors of the proposed triple selection based test (see Section \ref{sec:debiased}) are also close to the desired 5\% level of significance, but deviate in more extreme settings (see Figure \ref{fig:Setting1_1d} and \ref{fig:Setting1_1e}), namely when the effect of the covariates on censoring is rather weak (i.e., low values of $g$). This is somewhat surprising as we considered an additional model for censoring to improve the power of the linear regression test (see Step $2$ in Section \ref{sec:debiased}). In Figure 1 in Appendix B of the online supplementary materials, the results under a double selection approach (i.e., as described in Section \ref{sec:debiased} without the censoring model) show that the extra censoring model nonetheless has added value. 
Since the inflation of Type I errors is more pronounced in the scenario in Figure \ref{fig:Setting1_1d} than in the scenario in Figure \ref{fig:Setting1_1e}, a plausible reason might be that important predictors of the exposure are missed. 
We further observe that the poor man's approach and the proposed triple selection approach based on the de-biased test drastically outperform the proposal by \citet{fang2017} throughout large parts of the model space (see Figure \ref{fig:Setting1_1d} and \ref{fig:Setting1_1e}).

Based on the theoretical and Monte-Carlo results, we recommend the use of either the poor man's approach or triple selection over the estimator proposed by \citet{fang2017} and especially over the na\"ive post-Lasso estimator.
We refer to Figures 2-14 in Appendix B of the online supplementary materials for
results under additional settings. Figures 15-30 in Appendix B of the online supplementary materials show results under high-dimensional settings ($p>n$). Although none of the methods seems to maintain the nominal Type I error across the different settings, the poor man's approach outperforms the other approaches in the settings considered.

\section{Data Analysis}\label{sec:dataAnalysis}
We illustrate the proposal on the Breast cancer data set used in \cite{royston2013external}. In particular, we will investigate the effect of chemo on relapse free survival (in months). Data are available on 2,982 primary breast cancer patients whose data records were included in the Rotterdam tumor bank, including the following potential confounders: year of cancer incidence, age, menopausal status (pre- or post-menopausal), tumor size ($\leq 20$, 20-50, $>50$), tumor grade (2 or 3), number of positive lymph nodes, hormonal treatment (yes or no), progesterone receptors in fmol/l and estrogen receptors in fmol/l.

A standard Cox analysis adjusting for all covariates (using main effects) delivers a log hazard ratio of -0.12 (robust SE 0.07). Performing post-Lasso (with penalty parameter $\lambda_{1se}$ chosen via 20-fold cross-validation) results in a Cox model adjusted for tumor size, tumor grade and number of positive lymph nodes, and delivers a log hazard ratio of -0.02 (robust SE 0.07). The poor man's approach (with penalty parameter $\lambda_{1se}$ chosen via 20-fold cross-validation) additionally includes year of cancer incidence, age and menopausal status, and delivers a log hazard ratio of -0.13 (robust SE 0.07). Using the proposed triple selection approach, which in addition to the variables selected by Lasso includes year of cancer incidence, age, progesterone receptors and estrogen receptors, we obtain an estimated log hazard ratio of -0.14 (robust SE 0.07). 

Given the standard Cox analysis, the proposed estimators give results that look more plausible. To better understand the impact of the different approaches on resulting inferences, a sub-sampling approach was taken. Specifically, we take 1,000 subsamples of size 300 from the original dataset and evaluate how frequently the resulting 95\% confidence intervals contain the full-sample estimate obtained via a Cox model with all main effects include, which was taken as the benchmark. To construct confidence intervals, sandwich estimators of the standard errors were used (obtained via the robust option in Cox model-fitting software). We repeat this analysis using sub-samples of size 75, 150, 450 and 600.
Results in Table \ref{tab:dataAnalysis} suggest that the post-Lasso results in under-coverage and bias as a result of eliminating too many covariates. 
Better coverage and less biased results are consistently obtained by the poor man's and triple selection approach.

\begin{table}[ht]
	\centering
	\begin{tabular}{rlllll}
		\hline
		$n$ & Method & Bias & SD & Mean SE & Coverage \\ 
		\hline
		75 & Post-Lasso & 0.080 & 0.467 & 0.4145 & 0.915 \\ 
		& Poor man's approach & -0.020 & 0.550& 0.486 & 0.930 \\ 
		& Triple selection & -0.005 & 0.552 & 0.483 & 0.926 \\ 
		\hline
		150 & Post-Lasso & 0.088 & 0.317 & 0.288 & 0.916 \\ 
		& Poor man's approach & 0.001 & 0.355 & 0.328 & 0.938 \\ 
		& Triple selection & -0.008 & 0.359 & 0.328 & 0.939 \\ 
		\hline
		300 & Post-Lasso & 0.096& 0.208 & 0.204 & 0.910\\ 
		& Poor man's approach & 0.009& 0.234 & 0.229 & 0.952 \\ 
		& Triple selection & 0.002 & 0.235 & 0.228 & 0.948 \\ 
		\hline
		450 & Post-Lasso & 0.095 & 0.160 & 0.167 & 0.912 \\ 
		& Poor man's approach & 0.011 & 0.179 & 0.187 & 0.959 \\ 
		& Triple selection & 0.004 & 0.181 & 0.187& 0.951 \\ 
		\hline
		600 & Post-Lasso & 0.095 & 0.136 & 0.146 & 0.913 \\ 
		& Poor man's approach & 0.007 & 0.151 & 0.163 & 0.958 \\ 
		& Triple selection & 0.001 & 0.152 & 0.163 & 0.958 \\ 
		\hline
	\end{tabular}
	\caption{\label{tab:dataAnalysis} A comparision of the different variable selection methods. Bias: bias taken across subsamples, compared with the benchmark estimator; SD: standard deviation taken across sub-samples; mean SE: mean estimated standard error; coverage: coverage probability obtained via sub-sampling. 
	}
\end{table}

\section{Discussion}\label{sec:discussion}
In this work, we have developed simple to implement approaches for valid inference for conditional causal hazard ratios after variable selection. 
While both proposed tests perform well, the simpler poor man's approach emerged as the clear winner in Monte Carlo simulations. 
This has implications for other problems in causal inference where development and implementation of a ``debiased'' estimator  may be difficult (e.g., studies with time varying confounding).

Our triple selection estimator $\check{\alpha}$ for $\alpha^*$ is closely related to the one-step estimator proposed by \citet{fang2017}. These authors obtained a closed form decorrelated estimator
$$
\hat{\alpha}_\text{Lasso}-\left[\frac{\partial\mathbb{E}_n\left\{\hat{U}_i(\hat{\alpha}_\text{Lasso}, \hat{\beta}_\text{Lasso}, \hat{\gamma}_\text{Lasso} )\right\}}{\partial\alpha}\right]^{-1}\mathbb{E}_n\left\{\hat{U}_i(\hat{\alpha}_\text{Lasso}, \hat{\beta}_\text{Lasso}, \hat{\gamma}_\text{Lasso} )\right\}
$$
for $\alpha$ by linearizing (using a Newton-Raphson type correction) the estimating equation\\
$\mathbb{E}_n\left\{\hat{U}_i(\alpha, \hat{\beta}_\text{Lasso}, \hat{\gamma}_\text{Lasso} )\right\}=0,
$
at the initial estimator $\hat{\alpha}_\text{Lasso}$. Here, $\hat{\alpha}_\text{Lasso}$, $\hat{\beta}_\text{Lasso}$ and $\hat{\gamma}_\text{Lasso}$ are the initial Lasso estimators in Step $1$ and $3$ in Section \ref{sec:debiased}. 
Despite the fact that their method does not use a separate model for censoring, it is closely related to the triple selection estimator as it turns out that this estimator solves $\mathbb{E}_n\left\{\hat{U}_i(\alpha, \check{\beta}, \check{\gamma} )\right\}=0$ (see Appendix A of the online supplementary materials). Although the first order asymptotic properties of both estimators coincide, in finite samples, the triple selection method seems to deliver a more robust performance. This is likely due to the additional selection step for censoring and because the relevant gradients are likely made closer to zero within the sample by re-fitting the Cox model for survival.


A test based on the triple selection approach in Section \ref{sec:debiased} enjoys a specific form of double robustness under the null: if either the exposure mean is linear in $L$ or the log hazard of the survival endpoint is
linear in $L$ (but not necessarily both), then for similar reasons as in \cite{Dukes2020}, this test should be valid under the null when there is ultra-sparsity. This is both because the score function has mean zero if either model is correct, and because the proposed method for estimating $\beta^*$ and $\gamma^*$ ensures that the inference is doubly robust.
When choosing a working exposure model for the proposed de-biased test, we have focused on the identity link. 
To extend double robustness to the logit link (e.g., for a binary exposure), we could develop an analogous proposal starting from the following score function
\begin{align*}
    U_{i}\left(\alpha, \beta, \gamma, \gamma_0(t)\right)=\int_0^\tau\left\{A_{i}-\text{expit}\left(\gamma_0(t)+\gamma'L_i\right) \right\}\{dN_i(t)-R_i(t)\lambda_0(t, \alpha, \beta)e^{\alpha A_i+\beta' L_i}dt\}.
\end{align*}
We would then set the gradients with respect to $\gamma_0(t)$, $\gamma$ and $\beta$ approximately to zero by estimating the nuisance parameters as the solution to specific penalized equations given by those gradients.

Further, when choosing a working exposure model for the poor man's approach, we have mainly focused on the identity link. Other  link functions (e.g., logit) can be used in the three steps of the algorithm, so long as the corresponding penalized estimators of the nuisance parameters can be shown to converge sufficiently quickly.

The estimator on which our proposed test is based (along with common alternatives) has the disadvantage that it may not converge to something easily interpretable when the proportional hazards assumption fails. The resulting confidence intervals also typically lose their validity under misspecification of an assumed Cox proportional hazards model. This is not entirely satisfactory as there is no good understanding of what it infers in this case.
Inspired by developments on assumption-lean inference for generalized linear model parameters by \cite{Vansteelandt2020}, in future research we will develop nonparametric inference for a model-free estimand that reduces to a Cox model parameter under the proportional hazards assumption, but which continues to capture the association of interest under arbitrary types of misspecification.


\backmatter


\section*{Acknowledgements}

This project has received funding from Fulbright Belgium, Belgian American Educational Foundation, VLAIO under the Baekeland grant agreement HBC.2017.0219, BOF Grant BOF.01P08419 and FWO research projects G016116N and 1222522N.\\


%
  \bibliographystyle{biom} 
 \bibliography{mybibilo.bib}


\section*{Supporting Information}
Supplementary material is available at https://github.com/kelvlanc/TripleSelection.





\label{lastpage}

\end{document}